\documentclass[journal=jctcce, movefloats=false, layout=twocolumn]{achemso}
\usepackage[colorlinks=true,linkcolor=blue]{hyperref}
\usepackage{graphicx}
\usepackage{amsmath,amsfonts,amssymb}
\usepackage{color}
\usepackage{bm}
\usepackage[mathscr]{eucal}
\usepackage{picture}

\usepackage{booktabs}
\usepackage{multirow}
\usepackage{adjustbox}

\usepackage{graphicx}
\usepackage{amsmath,amsfonts,amssymb}
\usepackage{color}
\usepackage{bm}
\usepackage{threeparttable}
\usepackage[mathscr]{eucal}
\usepackage{picture}
\usepackage{threeparttable}
\usepackage{booktabs}
\usepackage{multirow}
\usepackage{adjustbox}

\definecolor{grey}{rgb}{0.7,0.7,0.7}

\usepackage{fancyvrb}
\usepackage{ulem}

\usepackage[T1]{fontenc}

\definecolor{brown}{RGB}{111,16,50}
\definecolor{purple}{rgb}{0.8,0.0,0.8}
\definecolor{pink}{rgb}{1.,0.5,0.5}

\newcommand{\ctext}[1]{\raise0.2ex\hbox{\textcircled{\scriptsize{#1}}}}

\title{TD$\Delta$SCF: Time-Dependent Density Functional Theory with a Non-Aufbau Reference for near-degenerate states}

\author{Shuto Shibasaki}
\affiliation{Department of Applied Chemistry, Shibaura Institute of Technology, 3-7-5 Toyosu, Koto-ku, Tokyo 135-8548 Japan}
\altaffiliation{These authors contributed equally to this work.}

\author{Fumiya Mohri}
\affiliation{Department of Applied Chemistry, Shibaura Institute of Technology, 3-7-5 Toyosu, Koto-ku, Tokyo 135-8548 Japan}
\altaffiliation{These authors contributed equally to this work.}
\alsoaffiliation{Present address: Graduate School of Science, Institute of Science Tokyo, 2-12-1 Ookayama, Meguro-ku, Tokyo 152-8551, Japan}
\author{Takashi Tsuchimochi}
\email{tsuchimochi@gmail.com}
\affiliation{College of Engineering, Shibaura Institute of Technology, 3-7-5 Toyosu, Koto-ku, Tokyo 135-8548 Japan}
\alsoaffiliation{Institute for Molecular Science, 38 Nishigonaka, Myodaiji, Okazaki 444-8585 Japan}
\begin{document}

\begin{abstract}
Near-degenerate electronic structures remain a major challenge for conventional single-reference density functional theory (DFT). To address this problem, we propose time-dependent $\Delta$SCF (TD$\Delta$SCF), a novel linear-response scheme in which a non-Aufbau $\Delta$SCF determinant serves as the reference for a subsequent TDDFT calculation. In contrast to collinear spin-flip (SF)-TDDFT, this formulation preserves the usual Coulomb and exchange-correlation response contributions while describing the target states from an electronically excited reference. We examine the performance of TD$\Delta$SCF for several prototypical problems involving near-degeneracy, including the torsional potential of ethylene, singlet--triplet gaps of representative diradicals, geometry optimizations of benzyne isomers, and bond-dissociation curves of hydrogen fluoride and F$_2$. Across these tests, TD$\Delta$SCF shows markedly weaker functional dependence than SF-TDDFT and often yields a more balanced description of challenging singlet states. In particular, it provides smooth torsional potentials, improved singlet--triplet gaps, a consistent monocyclic structure for singlet $m$-benzyne, and a more satisfactory description of bond dissociation without the spurious low-lying states found in SF-TDDFT. At the same time, the method exhibits a systematic tendency to overestimate singlet energies and can lose accuracy when the underlying $\Delta$SCF reference is not well suited to the final state. We also identify a numerical instability that can arise in non-Aufbau calculations and trace its origin to the exchange-correlation potential near uncompensated nodal regions. These results highlight both the promise and the practical limitations of TD$\Delta$SCF as a low-cost method for singlet states with near-degenerate electronic structures.

\end{abstract}
\maketitle

\section{Introduction}

Density functional theory (DFT) and its time-dependent extension (TDDFT) are among the most widely used electronic-structure methods because they offer a favorable balance between computational cost and accuracy.~\cite{KohnSham65, TDDFT, CasidaHuixRotllant} For many closed-shell molecules near their equilibrium geometries, conventional Kohn--Sham (KS) DFT and linear-response TDDFT provide practical descriptions of ground- and excited-state properties. However, these single-reference frameworks often become unreliable in the presence of orbital degeneracy or near-degeneracy, such as along bond-dissociation coordinates, near twisted geometries, and in diradical or polyradical systems. In such situations, the exact wave function acquires substantial multiconfigurational character, and a single-determinant reference is no longer qualitatively adequate.~\cite{Shao03,Krylov06}
 
A widely used low-cost strategy for treating such electronically degenerate cases is spin-flip time-dependent density functional theory (SF-TDDFT). In SF-TDDFT, low-spin target states are described as spin-flipping excitations from a high-spin reference determinant, allowing a formally single-reference response treatment to recover key features of near-degenerate electronic structure. This approach has been successfully applied to a broad range of chemical problems,  including diradicals,\cite{Shao03} conical intersections,\cite{Minezawa09} and photochemical dynamics.\cite{Harabuchi14, Yue18}

Despite these successes, collinear SF-TDDFT has a well-known formal limitation. Within the spin-flip manifold, the coupling is governed primarily by Hartree-Fock exchange, while the usual DFT exchange-correlation kernel does not contribute in the same way as in ordinary same-spin linear-response TDDFT. As a result, the accuracy of SF-TDDFT often shows pronounced functional dependence.~\cite{Li12,Bernard12} Although noncollinear SF-TDDFT restores the missing exchange-correlation coupling, its practical use is often hindered by numerical instability associated with the underlying kernel.~\cite{Huix10,Li12,Bernard12}

A different route to electronically excited states is provided by $\Delta$SCF approaches, in which the target excited configuration is treated through a variational orbital optimization of a non-Aufbau determinant, thereby incorporating state-specific orbital relaxation from the outset.
Since the introduction of practical excited-state SCF procedures such as the maximum overlap method (MOM),\cite{Gilbert08} such state-specific approaches have attracted renewed interest as computationally efficient orbital-optimized descriptions of excited states. In parallel with related developments such as restricted open-shell Kohn--Sham (ROKS) theory,\cite{Filatov99, Kowalczyk2013} $\Delta$SCF-based methods have been applied successfully to a wide range of excited states, and benchmark studies have shown that non-Aufbau state-specific descriptions can achieve accuracy comparable to, and in some cases better than, that of TDDFT.\cite{Hait20B, Hait21, Selenius2024, Paetow25} Their main limitation, however, is the difficulty of converging the desired excited-state solution reliably, because excited states often correspond to saddle points rather than minima on the orbital optimization landscape. To address this issue, several more robust optimization strategies have recently been developed, including the level-shift-based state-targeted energy projection (STEP) method,\cite{Carter-Fenk20} square gradient minimization (SGM),\cite{Hait20} and direct optimization approaches based on quasi-Newton or saddle-point-search algorithms.\cite{Levi20A, Levi20B} Together, these developments have substantially broadened the practical applicability of state-specific excited-state electronic structure methods.

From this perspective, a natural yet still largely unexplored next step is to formulate linear-response theory on top of a $\Delta$SCF excited-state reference. We refer to this approach as ``time-dependent $\Delta$SCF'' (TD$\Delta$SCF), which may also be viewed as non-Aufbau-reference TDDFT/TDA, to emphasize that a state-specific $\Delta$SCF reference is used as the starting point for a subsequent multi-state linear-response calculation. This idea is closely related to the physical picture underlying transient absorption spectroscopy (TAS), in which a system prepared by a pump pulse is interrogated by a subsequent probe pulse,\cite{Berera09} and it has very recently been exploited in simulating excited-state absorption relevant to TAS.\cite{Knepp25} Related concepts have also appeared in core-level spectroscopy, most notably in linear-response formalisms built on optimized hole references, such as the open-shell electron-attachment equation-of-motion coupled-cluster method of Nooijen and Bartlett\cite{Nooijen95} and electron-affinity TDDFT of Carter-Fenk and co-workers\cite{Carter-Fenk22}   for X-ray absorption. 
 These approaches share the broad philosophy of first incorporating important orbital-relaxation effects into a non-ground-state or hole reference and then performing a subsequent response/EOM calculation on top of that reference.

In the present work, however, our focus is different. Rather than targeting core-excited states or  excited-state absorption itself, we focus on the de-excitation channels relevant to {\it stimulated emission},\cite{Berera09} and on the ground-state-related information that can be extracted from them within such a TD$\Delta$SCF framework. This perspective parallels a wave function-based approach recently developed by one of us within the configuration-interaction-singles (CIS) framework, which showed encouraging performance for electronically degenerate systems.\cite{Tsuchimochi24,Tsuchimochi26A,Tsuchimochi26B} On this basis, we formulate TD$\Delta$SCF as a linear-response extension of $\Delta$SCF and assess its potential as a practical method for electronically challenging systems. In particular, TD$\Delta$SCF may offer an advantage over collinear SF-TDDFT because it naturally includes contributions from the exchange-correlation potential and Coulomb interaction that are absent from the collinear SF-TDDFT kernel and may become important for the property of interest. We also identify a previously unrecognized numerical instability that can arise in non-Aufbau calculations and elucidate its origin through an analysis of the associated exchange-correlation potential. Together, these results establish the formal and practical features of TD$\Delta$SCF while also providing new insight into the numerical behavior of $\Delta$SCF-based excited-state descriptions.

This paper is organized as follows. In Section II, we review the basic formulations of conventional TDDFT and SF-TDDFT and then introduce the theoretical framework of TD$\Delta$SCF, emphasizing its formal relation to and distinction from SF-TDDFT. Section III summarizes the computational details employed in the present study. In Section IV, we assess the performance of TD$\Delta$SCF through several representative applications, including the torsional potential of ethylene, singlet--triplet gaps of prototypical diradicals, geometry optimizations of the benzyne isomers, and bond-dissociation curves of hydrogen fluoride and F$_2$. We also examine a numerical instability that can arise in non-Aufbau calculations by analyzing the $\sigma_g \to \sigma_u$ excitation of H$_2$. Finally, Section V concludes the paper with a summary of the main findings and perspectives for future work.

\section{Theory}

\subsection{TDDFT}

In the density-matrix representation, the electron density $\rho({\bf x})$ is written as
\begin{equation}
\rho({\bf x})=\sum_{pq} P_{pq}\,\phi_p^*({\bf x})\phi_q({\bf x}),
\label{eq:rho_density_matrix}
\end{equation}
where ${\bf x}=(\mathbf r,\sigma)$ denotes the spatial and spin coordinates, and $\{\phi_p\}$ is a set of orthonormal spin-orbitals. For a single-determinantal $N$-electron reference, the one-particle density matrix $\mathbf P$ satisfies the idempotency condition
\begin{equation}
{\bf P}^2={\bf P},
\label{eq:idempotency}
\end{equation}
and its trace gives the total number of electrons.\cite{ParrYang,Ullrich}

The time evolution of the Kohn--Sham density matrix in the presence of a weak external perturbation is governed by
\begin{equation}
[{\bf F}+\lambda {\bf V}(t),{\bf P}]= i\frac{\partial {\bf P}}{\partial t},
\label{eq:tdks_density_matrix}
\end{equation}
where ${\bf V}(t)$ denotes an infinitesimal time-dependent field and $\mathbf F$ is the Kohn--Sham Fock matrix.\cite{TDDFT,Ullrich} In a spin-orbital basis, $\mathbf F$ is given by
\begin{equation}
\begin{aligned}
F_{pq}
={}&
\int \phi_p^*({\bf x})
\Biggl[
-\frac{1}{2}\nabla^2
-\sum_A \frac{Z_A}{|\mathbf r-\mathbf R_A|} \\
&\qquad
+\int \frac{\rho({\bf x}')}{|\mathbf r-\mathbf r'|}\,d{\bf x}'
+ v_{\mathrm{xc}}({\bf x})
\Biggr]
\phi_q({\bf x})\,d{\bf x}.
\end{aligned}
\label{eq:ks_fock}
\end{equation}

For weak perturbations, Eq.~\eqref{eq:tdks_density_matrix} may be linearized about the ground-state Kohn--Sham solution. Within the adiabatic approximation, the exchange-correlation potential at time $t$ depends only on the instantaneous density, so that the corresponding response kernel becomes frequency independent.\cite{TDDFT,CasidaHuixRotllant} In this linear-response regime, excitation energies appear as poles of the density-response function. When expressed in the basis of occupied--virtual orbital transitions, the resulting equations take the familiar Casida form,\cite{TDDFT,Petersilka96,Bauernschmitt96}
\begin{equation}
\begin{pmatrix}
{\bf A} & {\bf B}\\
{\bf B}^\ast & {\bf A}^\ast
\end{pmatrix}
\begin{pmatrix}
{\bf X}_I\\
{\bf Y}_I
\end{pmatrix}
=
\omega_I
\begin{pmatrix}
{\bf 1} & {\bf 0}\\
{\bf 0} & -{\bf 1}
\end{pmatrix}
\begin{pmatrix}
{\bf X}_I\\
{\bf Y}_I
\end{pmatrix},
\label{eq:casida}
\end{equation}
where $\omega_I$ is the excitation energy of state $I$, and ${\bf X}_I$ and ${\bf Y}_I$ are the forward and backward transition amplitudes, respectively. The matrix elements of ${\bf A}$ and ${\bf B}$ are given by
\begin{equation}
A_{ai,bj}
=
(\varepsilon_a-\varepsilon_i)\delta_{ab}\delta_{ij}
+
\frac{\partial F_{ai}}{\partial P_{bj}},
\label{eq:A_matrix}
\end{equation}
\begin{equation}
B_{ai,bj}
=
\frac{\partial F_{ai}}{\partial P_{jb}},
\label{eq:B_matrix}
\end{equation}
where indices $i,j,\ldots$ and $a,b,\ldots$ denote occupied and virtual orbitals, respectively.

Within the Tamm--Dancoff approximation (TDA), the coupling to the backward amplitudes is neglected,\cite{Hirata99} and Eq.~\eqref{eq:casida} reduces to
\begin{equation}
{\bf A}{\bf X}_I=\omega_I{\bf X}_I.
\label{eq:tda}
\end{equation}
Thus, TDDFT/TDA may be viewed as a CIS-like eigenvalue problem in which the bare Kohn--Sham orbital-energy differences are supplemented by Coulomb and exchange-correlation response contributions.\cite{Hirata99,CasidaHuixRotllant}

The limitations of conventional TDDFT for near-degenerate systems can be understood directly from Eq.~\eqref{eq:tda}. The response space is constructed from spin-conserving one-electron excitations based on a single ground-state Kohn--Sham determinant. When the physically relevant low-energy states require more than one dominant configuration, as in bond breaking or diradicals, the reference determinant itself becomes qualitatively unbalanced. In such cases, the difficulty originates from the choice of reference state and the associated response manifold.\cite{CasidaHuixRotllant,Krylov06} This observation motivates alternative response formalisms in which low-spin target states are accessed from a more balanced high-spin reference through spin-flip excitations.\cite{Shao03,Krylov06}

\subsection{SF-TDDFT}

As in other spin-flip methods,\cite{Casanova20} SF-TDDFT replaces the ordinary low-spin reference with a high-spin determinant, typically an $M_S=1$ triplet reference, and solves Eq.~\eqref{eq:tda} in the spin-flip excitation space. The corresponding excitation operator is
\begin{equation}
\hat E_{i_\alpha}^{a_\beta}
=
a_{a_\beta}^{\dagger} a_{i_\alpha},
\label{eq:sf_operator}
\end{equation}
which promotes an electron from an occupied $\alpha$ orbital to a virtual $\beta$ orbital while simultaneously flipping its spin.

The matrix elements in the spin-flip block are given by
\begin{equation}
A^{\mathrm{SF}}_{ a_\beta i_\alpha,b_\beta j_\alpha }
=
(\varepsilon_{a_\beta}-\varepsilon_{i_\alpha})\delta_{ij}\delta_{ab}
+
K^{\mathrm{SF}}_{a_\beta i_\alpha,b_\beta j_\alpha }.
\label{eq:A_sf}
\end{equation}
In the collinear formulation, however, this coupling term is qualitatively different from that in ordinary TDDFT.
In particular, the Coulomb term does not couple spin-flip excitations,
\begin{equation}
\frac{\partial J_{a_\beta i_\alpha}}{\partial P_{b_\beta j_\alpha }} = 0,
\label{eq:coulomb_zero_sf}
\end{equation}
and the semilocal exchange-correlation kernel does not provide the usual off-diagonal coupling in this block. As a result, the dominant nontrivial contribution is given by the Hartree--Fock exchange part, so that Eq.~\eqref{eq:A_sf} is effectively reduced to
\begin{equation}
A^{\mathrm{SF}}_{a_\beta i_\alpha ,b_\beta j_\alpha }
= 
(\varepsilon_{a_\beta}-\varepsilon_{i_\alpha})\delta_{ij}\delta_{ab}
-
c_{\mathrm{HF}}
( a_\beta b_\beta | i_\alpha j_\alpha ),
\label{eq:A_sf_hf}
\end{equation}
where $c_{\mathrm{HF}}$ is the fraction of Hartree--Fock exchange.

Equation~\eqref{eq:A_sf_hf} shows that, within collinear SF-TDDFT, the coupling between spin-flip configurations is governed primarily by the Hartree--Fock exchange term. This is the origin of the pronounced functional dependence of collinear SF-TDDFT and motivates the search for an alternative formulation that retains the exchange-correlation response more completely. To address this deficiency, Wang and Ziegler introduced noncollinear SF-TDDFT, in which the exchange-correlation response is formulated for noncollinear spin densities and can therefore include the transverse spin contributions missing in the collinear treatment.\cite{Wang04} This noncollinear formulation can substantially reduce the strong functional dependence of collinear SF-TDDFT.\cite{Wang05,Bernard12} In practice, however, noncollinear kernels beyond the local-density approximation are more difficult to construct and may suffer from numerical instabilities, particularly for generalized gradient approximation (GGA) functionals.\cite{Li12,Bernard12} For this reason, adiabatic local-density-type kernels are often employed in practical calculations.\cite{Li12}

\subsection{TD$\Delta$SCF}
\begin{figure}[t]
\centering
\includegraphics[width=0.48\textwidth]{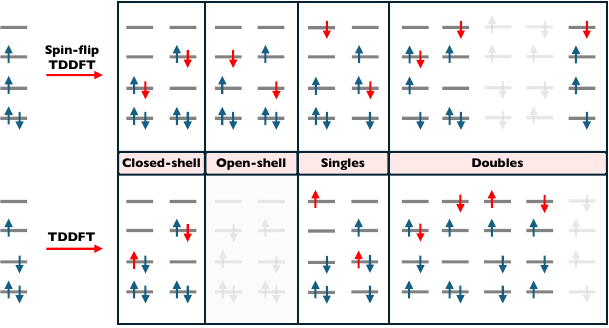}
\caption{Illustration of SF-TDDFT and TD$\Delta$SCF. The two methods employ different reference states. SF-TDDFT uses a high-spin configuration and accesses low-spin states via spin-flip excitations. In TD$\Delta$SCF, the reference is an excited-state configuration and the target states are described by subsequent linear-response excitations. Grayed-out configurations are not accessible within each linear-response space.}
\label{fig:Illustration_some_method}
\end{figure}

Another, less explored route for treating near-degenerate electronic structures within a single-determinant framework is to use a non-Aufbau excited-state determinant, as in $\Delta$SCF, as the reference for a subsequent linear-response treatment. 
We introduce TD$\Delta$SCF as a new linear-response framework in which a non-Aufbau $\Delta$SCF determinant serves as the reference state for a subsequent TDDFT/TDA calculation. The key idea is to retain the standard linear-response structure of TDDFT while replacing the ground-state Kohn--Sham reference by an electronically promoted self-consistent determinant. A schematic comparison of TD$\Delta$SCF with SF-TDDFT is shown in Fig.~\ref{fig:Illustration_some_method}.

The TDA equation retains the same algebraic form as Eq.~\eqref{eq:tda},
\begin{equation}
\mathbf A^{\Delta\mathrm{SCF}}
\mathbf X_I^{\Delta\mathrm{SCF}}
=
\omega_I^{\Delta\mathrm{SCF}}
\mathbf X_I^{\Delta\mathrm{SCF}},
\label{eq:tda_td_dscf}
\end{equation}
but the matrix is now defined with respect to the $\Delta$SCF reference determinant $|\Phi_{\Delta\mathrm{SCF}}\rangle$:
\begin{equation}
A^{\Delta\mathrm{SCF}}_{ai,bj}
=
(\varepsilon_a^{\Delta\mathrm{SCF}}-\varepsilon_i^{\Delta\mathrm{SCF}})
\delta_{ij}\delta_{ab}
+
K^{\Delta\mathrm{SCF}}_{ai,bj}.
\label{eq:A_td_dscf}
\end{equation}

The corresponding states are expressed as
\begin{equation}
|\Psi_I^{\Delta\mathrm{SCF}}\rangle
\sim
\sum_{ai}
X_{ai}^{I,\Delta\mathrm{SCF}}
\hat E_i^a
|\Phi^{\Delta\mathrm{SCF}}\rangle,
\label{eq:state_td_dscf}
\end{equation}
which may be viewed as response states built on the promoted reference and, in favorable cases, can describe states with de-excitation-like character relative to $|\Phi^{\Delta\mathrm{SCF}}\rangle$.

The essential difference from SF-TDDFT is that TD$\Delta$SCF changes the reference determinant itself while retaining a spin-conserving response manifold. As a result, the usual Coulomb and exchange-correlation response terms are retained in the kernel $K^{(\Delta\mathrm{SCF})}_{ai,bj}$ of Eq.~\eqref{eq:A_td_dscf}, unlike in the collinear spin-flip block, where the coupling is effectively dominated by the Hartree--Fock exchange contribution. In this way, TD$\Delta$SCF is intended to provide a reference better suited to a promoted electronic configuration without sacrificing the standard exchange-correlation response structure of conventional TDDFT.

TD$\Delta$SCF may therefore offer a promising route to a more accurate and less functional-sensitive description of near-degenerate electronic structures, although this expectation must ultimately be assessed numerically. In particular, because the response problem remains spin-conserving, its functional dependence may be closer to that of ordinary TDDFT than to that of collinear SF-TDDFT.
A further practical advantage of TD$\Delta$SCF is its simplicity of implementation. Once a suitable $\Delta$SCF reference has been obtained, the subsequent calculation formally reduces to an ordinary spin-conserving TDDFT/TDA step. Thus, in existing electronic-structure codes that already support both $\Delta$SCF and TDDFT, TD$\Delta$SCF can in principle be realized with only minimal additional implementation, making it straightforward to test in practice.

As in SF-TDDFT, however, the choice of reference determinant is critical: the non-Aufbau reference must be constructed so that, after linear response, it properly represents the near-degeneracy of the system under study. In practice, this requires a determinant built from the appropriate open-shell orbitals; otherwise, the resulting response states may be physically meaningless. Another practical challenge is that $\Delta$SCF calculations are often difficult to converge, because the SCF optimization is prone to variational collapse back to the ground-state determinant. Nevertheless, a number of algorithms have been developed to stabilize convergence to the desired $\Delta$SCF solution, including MOM,\cite{Gilbert08} STEP,\cite{Carter-Fenk20} and SGM.\cite{Hait20}

We should also note that TD$\Delta$SCF shares many configurations in common with SF-TDDFT as illustrated schematically in Fig.~\ref{fig:Illustration_some_method},  the two methods do not possess the same response space. A chief difference is that, in TD$\Delta$SCF, the open-shell configuration itself is used as the reference determinant; therefore, the open-shell singlet and triplet states that appear as spin-flip solutions in SF-TDDFT are not obtained as response states within the same TD$\Delta$SCF calculation and are consequently grayed out in Fig.~\ref{fig:Illustration_some_method}.  
The remaining response configurations are nevertheless closely related to those generated in SF-TDDFT: they are either the same configurations or the spin-complementary configurations obtained by interchanging the spin labels of the corresponding spin-flip configurations. 
In this sense, TD$\Delta$SCF and SF-TDDFT span related configuration manifolds, but differ in which configuration is chosen as the reference and which configurations appear as response states.

This relation also implies that TD$\Delta$SCF suffers from

the same spin-contamination issue as SF-TDDFT. Because the singly excited non-Aufbau $\Delta$SCF state typically used in TD$\Delta$SCF is a mixture of singlet and triplet configurations, the resulting TDA solutions are often spin contaminated. In practice, however, as will be demonstrated, the lowest states (obtained mainly as a linear combination of closed-shell configurations in Fig.~\ref{fig:Illustration_some_method}) exhibit only very small spin contamination, similar to what is observed in SF-TDDFT. 
Recent mixed-reference extensions of SF-TDDFT by Choi and co-workers have addressed related spin-contamination and spin-adaptation issues in the spin-flip framework.\cite{MRSFTDDFT} 
Although the present work does not employ such a mixed-reference construction, similar ideas may be useful for further reducing spin contamination in TD$\Delta$SCF.

Another important difference between TD$\Delta$SCF and SF-TDDFT is that the former cannot, in principle, describe triplet states within the same response space, whereas the latter includes them naturally among its solutions. Consequently, triplet states must be treated in separate calculations within the TD$\Delta$SCF framework.

\section{Computational details}
For TD$\Delta$SCF calculations, we first optimized the ground-state Kohn--Sham determinant. Starting from the resulting Kohn--Sham orbitals, we then constructed the desired $\Delta$SCF reference state by promoting an $\alpha$ electron from an occupied orbital to a virtual orbital so as to generate the appropriate open-shell electronic configuration. In many cases, this corresponds to a bonding-to-antibonding excitation. To stabilize convergence of the $\Delta$SCF procedure, we employed the maximum-overlap method (MOM). For comparison, we also carried out collinear SF-TDDFT calculations using the same open-shell configuration as the reference. Noncollinear SF-TDDFT was not considered in the present study; a more comprehensive assessment of that approach is left for future work. For benchmarking, we primarily employed the BLYP, B3LYP, and BHHLYP functionals\cite{B3LYP} to assess the effect of the Hartree--Fock exchange fraction, with $c_{\rm HF} = 0$, 0.2, and 0.5, respectively. We emphasize that low- and moderate-exchange functionals such as BLYP and B3LYP are not the standard recommended choices for collinear SF-TDDFT, for which functionals with a large fraction of Hartree--Fock exchange are usually preferred.\cite{Casanova20} The SF-BLYP and SF-B3LYP calculations are therefore included only as diagnostic comparisons to examine the functional dependence of the spin-flip kernel under a controlled variation of the Hartree--Fock exchange fraction, not as recommended SF-TDDFT protocols. For the conventional DFT calculations used for comparison, the singlet states were computed using restricted Kohn--Sham calculations, whereas the triplet states were obtained from unrestricted Kohn--Sham calculations. 

 All calculations were performed with Q-Chem,\cite{QChem} except for the grid analysis in Section~\ref{sec:grid}, which was performed using PySCF.\cite{pyscf}

For the singlet--triplet gap calculations discussed in Section~\ref{sec:st_gap}, the triplet states in TD$\Delta$SCF were obtained from separate unrestricted KS calculations, whereas in SF-TDDFT they were taken as low-spin triplet solutions of the TDA equation.

\begin{figure}
\raggedright
\includegraphics[width=0.5\textwidth]{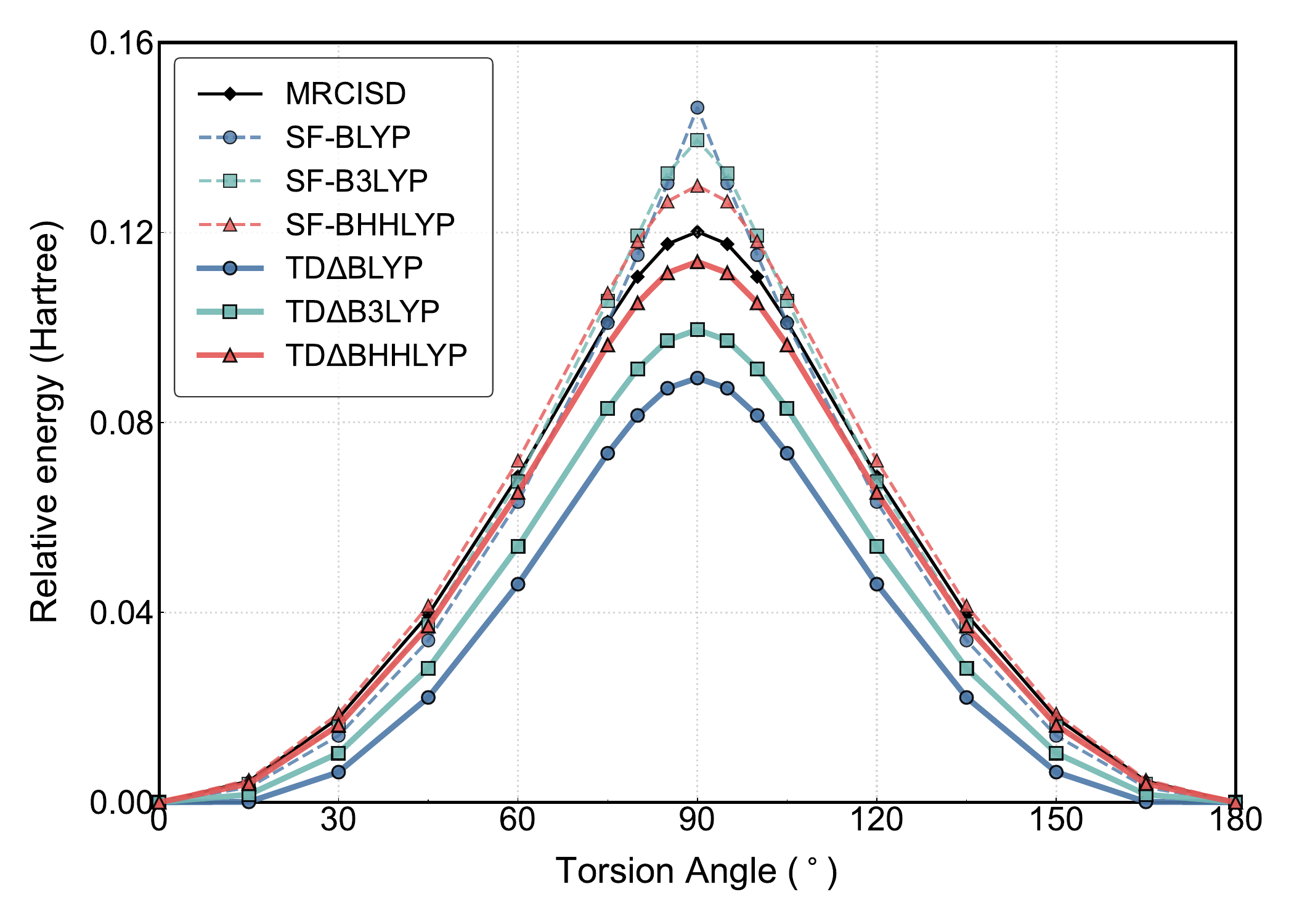}
\caption{Torsional potential energy curves of ethylene computed with TD$\Delta$SCF (solid line), SF-TDDFT (dashed line) and MRCISD using the DZP basis set. A cusp is clearly observed near the barrier region for the pure functional (BLYP). Energies are referenced to the planar (0$^\circ$) geometry.}
\label{fig:ethylene_torsion}
\end{figure}

\begin{table*}[htbp]
\centering
\footnotesize\setlength{\tabcolsep}{1.5pt}
\caption{Torsion-angle-dependent singlet energies $E$ and spin expectation values $\langle S^2\rangle$ of ethylene computed with TD$\Delta$SCF and SF-TDDFT using the BLYP, B3LYP, and BHHLYP functionals. The unrelaxed torsional barrier height $\Delta E = E(90^\circ)-E(0^\circ)$ (eV) is also reported. }
\label{tb:torsion_energy_S2_col}
\begin{tabular}{lccccccccccccccccc}
\hline\hline
& \multicolumn{8}{c}{TD$\Delta$SCF} && \multicolumn{8}{c}{SF-TDDFT} \\
\cline{2-9}\cline{11-18}
Angle
& \multicolumn{2}{c}{BLYP} && \multicolumn{2}{c}{B3LYP} && \multicolumn{2}{c}{BHHLYP}
&& \multicolumn{2}{c}{BLYP} && \multicolumn{2}{c}{B3LYP} && \multicolumn{2}{c}{BHHLYP} \\
\cline{2-3}\cline{5-6}\cline{8-9}\cline{11-12}\cline{14-15}\cline{17-18}
& $E$ & $\langle S^2\rangle$ && $E$ & $\langle S^2\rangle$ && $E$ & $\langle S^2\rangle$
&& $E$ & $\langle S^2\rangle$ && $E$ & $\langle S^2\rangle$ && $E$ & $\langle S^2\rangle$ \\
\hline
$0^\circ$  & -78.50429 & 0.0123 && -78.56900 & 0.0122 && -78.53267 & 0.0128 && -78.49553 & 0.0085 && -78.56037 & 0.0099 && -78.52842 & 0.0140 \\
$15^\circ$ & -78.50421 & 0.0138 && -78.56736 & 0.0135 && -78.52876 & 0.0145 && -78.49239 & 0.0085 && -78.55649 & 0.0100 && -78.52369 & 0.0142 \\
$30^\circ$ & -78.49794 & 0.0148 && -78.55864 & 0.0147 && -78.51645 & 0.0166 && -78.48150 & 0.0084 && -78.54424 & 0.0101 && -78.50975 & 0.0147 \\
$45^\circ$ & -78.48220 & 0.0149 && -78.54079 & 0.0153 && -78.49556 & 0.0183 && -78.46143 & 0.0085 && -78.52293 & 0.0104 && -78.48703 & 0.0155 \\
$60^\circ$ & -78.45836 & 0.0144 && -78.51514 & 0.0157 && -78.46742 & 0.0200 && -78.43226 & 0.0090 && -78.49275 & 0.0111 && -78.45647 & 0.0163 \\
$75^\circ$ & -78.43079 & 0.0126 && -78.48608 & 0.0153 && -78.43635 & 0.0220 && -78.39459 & 0.0101 && -78.45474 & 0.0118 && -78.42117 & 0.0159 \\
$80^\circ$ & -78.42282 & 0.0114 && -78.47774 & 0.0149 && -78.42750 & 0.0227 && -78.38028 & 0.0106 && -78.44098 & 0.0117 && -78.41025 & 0.0147 \\
$85^\circ$ & -78.41709 & 0.0102 && -78.47172 & 0.0144 && -78.42119 & 0.0231 && -78.36516 & 0.0112 && -78.42795 & 0.0107 && -78.40188 & 0.0131 \\
$90^\circ$ & -78.41495 & 0.0097 && -78.46948 & 0.0142 && -78.41884 & 0.0233 && -78.34924 & 0.0551 && -78.42092 & 0.0089 && -78.39857 & 0.0122 \\
\hline
$\Delta E$ [eV]
& 2.43 &  && 2.71 &  && 3.10 &  & & 3.98 &  && 3.79 &  && 3.53 &  \\
\hline\hline
\end{tabular}
\end{table*}

\section{Results}

\subsection{Ethylene torsion}\label{sec:ethylene}
The torsional potential energy curve of ethylene has been widely used as a benchmark for discussing multiconfigurational character.~\cite{Shao03,Li12,Saade24} In the planar geometry, ethylene forms a $\pi$ bond; however, as the torsion angle increases, the $\pi$ bond progressively breaks, and a proper description requires a superposition of two configurations corresponding to the $\pi$  and $\pi^*$  orbitals. Consequently, single-reference methods such as conventional DFT and Hartree-Fock become inadequate and often lead to a nonphysical cusp in the potential energy curve.

The ethylene geometry used in the calculations was taken from~\cite{Krylov01_B}, with structural parameters fixed at $r_{\rm CC}=1.330$~{\AA} , $r_{\rm CH}=1.076$~{\AA}, and $d_{\rm HCH}=116^\circ$. We have employed the DZP basis set. Taking the singlet energy at the planar structure (0$^\circ$ torsion) as the reference, we evaluated the torsion-induced energy differences and plotted the resulting curves in Figure ~\ref{fig:ethylene_torsion}. The corresponding singlet energies and spin expectation values, $\langle S^2 \rangle$, are summarized in Table~\ref{tb:torsion_energy_S2_col}.

\begin{table*}[htbp]
\tabcolsep=10pt
\centering
\small
\setlength{\tabcolsep}{3.5pt}
\caption{ Ground-state reference energies at $0^\circ$ (Hartree), deviations of TD$\Delta$SCF and SF-TDDFT from standard KS-DFT at $0^\circ$ (mHartree), and torsional barrier heights of ethylene defined as $\Delta E_X = E_X(90^\circ)-E_{\mathrm{KS-DFT}}(0^\circ)$ (eV), where $X=$ TD$\Delta$SCF or SF-TDDFT.}
\label{tb:torsion_energy0_90}
\begin{tabular}{lrrrrrr}
 \hline\hline
 Functional & $E_{\rm KS-DFT }(0^\circ)$ & TD$\Delta$SCF  (mH)& SF-TDDFT (mH) & $\Delta E_{\mathrm{TD}\Delta\mathrm{SCF}}$ (eV) & $\Delta E_{\mathrm{SF-TDDFT}}$ (eV)\\
 \hline
 BLYP   & -78.53863 & 34.3 & 43.1 & 3.37 & 5.15 \\
 B3LYP  & -78.59057 & 21.6 & 30.2 & 3.30 & 4.62 \\
 BHHLYP & -78.53476 & 2.1  & 6.3  & 3.15 & 3.71 \\
 \hline\hline
\end{tabular}
\end{table*}

All TD$\Delta$SCF methods tested, i.e., TD$\Delta$BLYP, TD$\Delta$B3LYP, and TD$\Delta$BHHLYP, produced smooth potential energy curves in the barrier region around 90$^\circ$. In contrast, the SF-BLYP curve exhibits a cusp, which can be attributed to the absence of coupling between distinct spin-flip excitations in pure exchange-correlation functionals. The computed torsional barriers were 3.10 and 3.53 eV obtained with TD$\Delta$BHHLYP and SF-BHHLYP, respectively, both in good agreement with the high-level multireference value of 3.27eV obtained by MRCISD~\cite{Nikiforov14}. Thus, the near-degenerate region around $90^\circ$ is described reasonably well by TD$\Delta$SCF regardless of the functional.

However, TD$\Delta$SCF tends to underestimate the barrier height when BLYP and B3LYP are used. TD$\Delta$BLYP and TD$\Delta$B3LYP yield barrier heights of 2.43 and 2.71 eV, corresponding to errors of $-0.84$ and $-0.56$ eV relative to the MRCISD value, respectively. By contrast, SF-TDDFT overestimates the barriers, giving 3.98 and 3.79 eV with BLYP and B3LYP, respectively, corresponding to errors of $+0.71$ and $+0.52$ eV. This underestimation in TD$\Delta$SCF originates from the deteriorated description at $0^\circ$. Since the $0^\circ$ structure is not degenerate, it can be described accurately by a standard  ground-state DFT calculation. Table~\ref{tb:torsion_energy0_90} summarizes the energies at 0$^\circ$ computed with standard ground-state DFT $E_{\rm KS-DFT}(0^\circ)$ for each functional. We find that, as the HF exchange fraction decreases, both the TD$\Delta$SCF and SF-TDDFT energies deviate further from $E_{\rm KS-DFT}(0^\circ)$. We also report in Table~\ref{tb:torsion_energy0_90} the barrier heights evaluated using the KS-DFT reference energy, $\Delta E_{\mathrm{TD}\Delta\mathrm{SCF}} =E_{\mathrm{TD}\Delta\mathrm{SCF}}(90^\circ)-E_{\mathrm{KS-DFT}}(0^\circ)$. When this energy difference is used, reasonable agreement not only with MRCISD but also among different functionals is obtained: the resulting barriers are 3.37, 3.30, and 3.15 eV for BLYP, B3LYP, and BHHLYP, respectively.

\begin{figure*}
    \centering
    \includegraphics[width=0.95\linewidth]{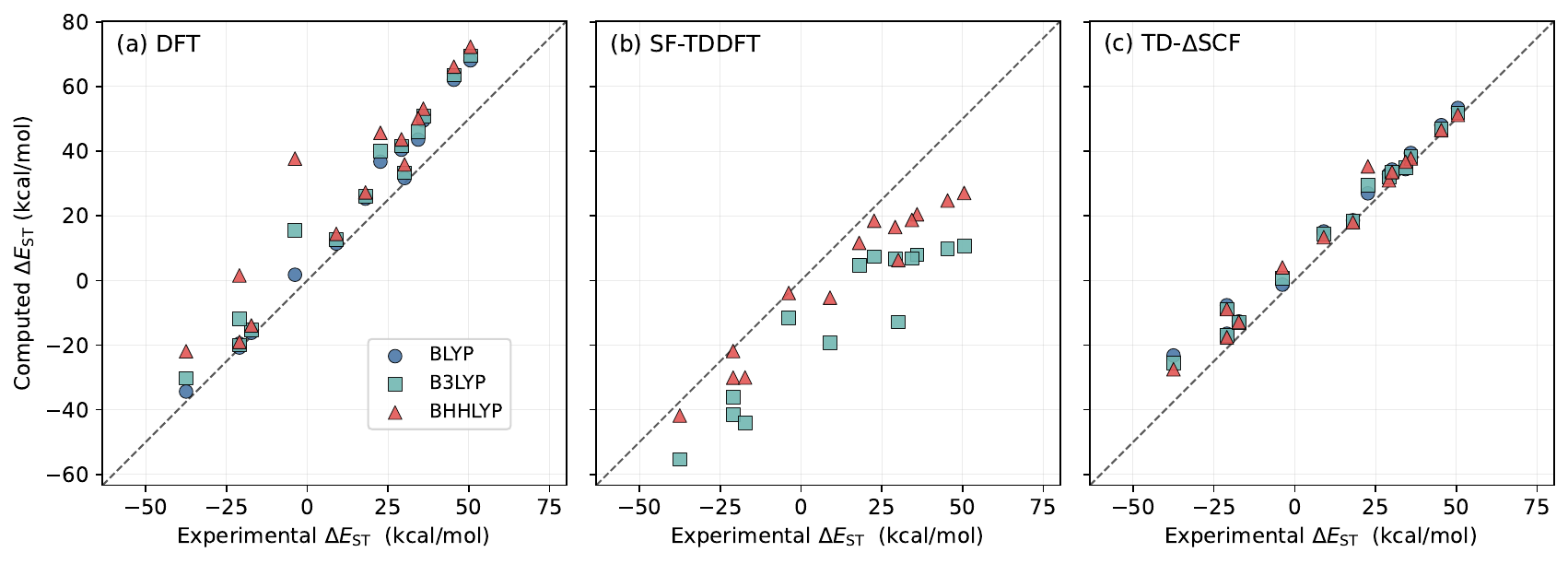}
    \caption{Comparison of computed and experimental $\Delta E_{\rm ST}$ values (kcal/mol) obtained with different methods. For SF-TDDFT, the BLYP results are omitted; see the main text for details.}
    \label{fig:st_gap}
\end{figure*}

\begin{table*}
\small
\setlength{\tabcolsep}{1.5pt}
\caption{Errors in the calculated $\Delta E_{\rm ST}$ values relative to experiment (kcal/mol) for each method and functional.}
\label{tb:st_gap}
\begin{threeparttable}
\begin{tabular}{l c c c c c c c c c c c c}
\hline\hline
System
& \multicolumn{3}{c}{DFT}
& & \multicolumn{3}{c}{SF-TDDFT}
& & \multicolumn{3}{c}{TD$\Delta$SCF}
& Reference$^{a}$ \\ \cline{2-4} \cline{6-8} \cline{10-12}
& BLYP & B3LYP & BHHLYP
& & BLYP & B3LYP & BHHLYP
& & BLYP & B3LYP & BHHLYP
& \\ \hline
C              & 11.30 & 12.46 & 14.50 & & -191.31 & -22.38 & -12.62 & & 3.77  & 2.75  & 1.84  & 29.15 \\
O              & 16.74 & 18.17 & 20.79 & & -202.38 & -35.60 & -20.63 & & 2.68  & 1.52  & 1.13  & 45.36 \\
Si             & 7.31  & 8.02  & 9.23  & & -152.82 & -13.30 & -6.38  & & 0.62  & 0.30  & 0.03  & 18.01 \\
NH             & 13.57 & 14.91 & 17.21 & & -173.22 & -27.97 & -15.45 & & 3.50  & 2.47  & 1.82  & 35.93 \\
OH$^+$         & 17.59 & 19.01 & 21.79 & & -305.84 & -39.86 & -23.42 & & 2.84  & 1.35  & 0.74  & 50.50 \\
O$_2$          & 14.12 & 17.48 & 23.04 & & -317.83 & -15.26 & -4.16  & & 4.31  & 6.92  & 12.65 & 22.64 \\
NF             & 9.26  & 11.63 & 15.83 & & -272.75 & -27.49 & -15.58 & & 0.12  & 0.62  & 2.43  & 34.32 \\
CH$_2$         & 2.43  & 3.77  & 5.44  & & -224.83 & -28.26 & -14.31 & & 6.18  & 5.48  & 4.40  & 8.99  \\
NH$_2^+$       & 1.56  & 3.19  & 5.78  & & -349.85 & -42.98 & -23.82 & & 4.18  & 3.51  & 3.40  & 30.12 \\
SiH$_2$        & 0.24  & 0.98  & 1.98  & & -194.21 & -20.46 & -9.07  & & 4.51  & 4.04  & 3.40  & -20.98 \\
PH$_2^+$       & 1.11  & 1.94  & 3.39  & & -254.34 & -26.71 & -12.70 & & 4.62  & 4.28  & 4.47  & -17.30 \\
$o$-benzyne    & 3.14  & 7.23  & 15.59 & & -98.11  & -17.85 & -4.28  & & 14.30 & 11.93 & 10.04 & -37.5 \\
$m$-benzyne    & 1.46  & 9.08  & 22.56 & & -98.33  & -15.11 & -0.84  & & 13.34 & 12.17 & 12.13 & -21.0 \\
$p$-benzyne    & 5.57  & 19.31 & 41.51 & & -89.05  & -7.69  & -0.08  & & 2.50  & 4.43  & 7.84  & -3.8 \\
ME             & 7.53  & 10.51 & 15.62 & & -208.92 & -24.35 & -11.67 & & 4.82  & 4.41  & 4.74  &        \\
MAE            & 7.53  & 10.51 & 15.62 & & 208.92  & 24.35  & 11.67 & & 4.82  & 4.41  & 4.74  &        \\ 
\hline\hline
\end{tabular}
\begin{tablenotes}[flushleft]
\footnotesize
\item[$^{a}$] Experimental values for the atoms were taken from Ref.~\citenum{atoms}, those for the diatomic molecules were taken from Ref.~\citenum{diatomic}. The reference values for CH$_2$, NH$_2^+$, SiH$_2$, and PH$_2^+$ were taken from Refs.~\citenum{CH2,NH2+,SiH2,PH2+}, respectively. The reference values for the benzyne isomers were taken from Ref.~\citenum{benzyne}.
\end{tablenotes}
\end{threeparttable}
\end{table*}

This analysis suggests that, in TD$\Delta$SCF, the near-degenerate region around 90$^\circ$ is described reasonably well regardless of the functional, whereas near 0$^\circ$, the accuracy deteriorates as the fraction of Hartree-Fock exchange decreases. In SF-TDDFT, by contrast, the description deteriorates at both 0$^\circ$ and 90$^\circ$. This is reflected in the corresponding barrier heights evaluated with the same KS-DFT reference, which are 5.15, 4.62, and 3.71 eV for BLYP, B3LYP, and BHHLYP, respectively. Of course, SF-BLYP does not give a correct picture at 90$^\circ$, so  the apparently reasonable barrier height in the original definition is due to fortuitous error cancellation  arising from overestimation at both 0$^\circ$ and 90$^\circ$.

Finally, the spin expectation values $\langle S^2 \rangle$ indicate that TD$\Delta$SCF, similar to SF-TDDFT, exhibits sufficiently small spin contamination (less than 0.1 ) and yields physically reasonable results.

\subsection{Singlet-triplet energy gaps} \label{sec:st_gap}
To further quantitatively assess the performance of TD$\Delta$SCF, we  next evaluate the singlet--triplet energy gaps $\Delta E_{\rm ST}$ of representative diradicals and compare them with those obtained by standard KS-DFT and SF-TDDFT. 
To this end, we consider the following systems: (1) the C, O, and Si atoms; (2) the diatomic molecules NH, OH$^+$, O$_2$, and NF; (3) the carbene derivatives CH$_2$, NH$_2^+$, SiH$_2$, and PH$_2^+$; and (4) the $o$-, $m$-, and $p$-benzyne isomers. For CH$_2$, the geometry optimized at the FCI/TZ2P level was used, whereas those for the carbene derivatives were taken from CISD/TZ2P(f,d) optimizations. For the benzyne systems, the geometries optimized at the SF-TDDFT/50-50(50\% Hartree--Fock + 8\% Slater + 42\% Becke for exchange and 19\% VWN + 81\% LYP for correlation) /6-311G* level were used. The geometries of the carbene and benzyne systems were the same as those used in Ref.\cite{Bernard12}. In addition, the geometries of the diatomic molecules were optimized at the MP2/TZ2P level. These systems provide a stringent test because conventional DFT describes only a single determinantal state and is therefore often inadequate for near-degenerate electronic structures.

\begin{figure*}
    \centering
    \includegraphics[width=0.95\linewidth]{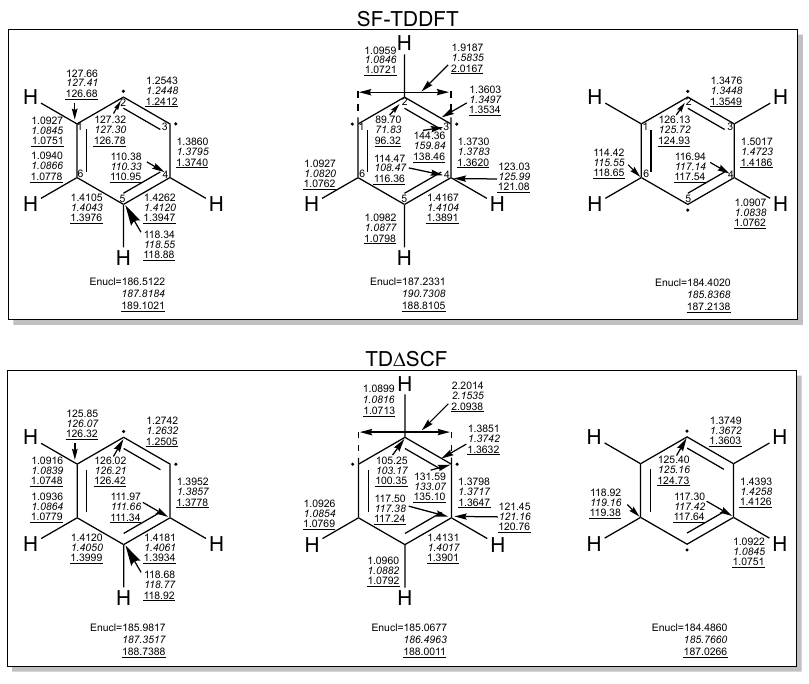}
    \caption{Optimized singlet-state geometrical parameters of the benzyne systems obtained by SF-TDDFT and TD-$\Delta$SCF, together with the nuclear repulsion energies (bond lengths in \AA, bond angles in degrees, and energies in hartree). BLYP is indicated in normal type, B3LYP in italics, and BHHLYP and 50-50 by underlining.}
    \label{fig:geom}
\end{figure*}

The overall trends are illustrated in Fig.~\ref{fig:st_gap}, where the calculated $\Delta E_{\rm ST}$ are plotted against the experimental values, taken from Ref.\cite{atoms,diatomic,CH2,NH2+,SiH2,PH2+,benzyne}.  Conventional DFT shows substantial deviations from experiment, and these deviations increase as the Hartree--Fock exchange fraction increases. This trend is quantified in Table~\ref{tb:st_gap}, which summarizes the signed errors relative to experiment. The MAEs of conventional DFT are 7.53, 10.51, and 15.62 kcal/mol for BLYP, B3LYP, and BHHLYP,  respectively. SF-TDDFT performs especially poorly with BLYP, for which the results deviate catastrophically from the experimental values, giving an MAE of 208.92 kcal/mol. For this reason, the SF-BLYP results are omitted from Fig.~\ref{fig:st_gap}. Although the results improve substantially when Hartree--Fock exchange is introduced, the scatter remains large even with B3LYP and BHHLYP, particularly for atoms, small diatomic molecules, and carbenes with pronounced one-center diradical character, for which limitations of collinear SF-TDDFT have been noted previously.\cite{Bernard12}  In addition, SF-TDDFT systematically underestimates the gaps. Moreover, relative to conventional DFT, SF-TDDFT with B3LYP is actually less accurate overall, and even with BHHLYP the improvement is only marginal. These results clearly demonstrate that SF-TDDFT exhibits a much stronger functional dependence than conventional DFT.

In contrast, TD$\Delta$SCF can incorporate exchange--correlation effects that are not properly captured by SF-TDDFT. As a result, it avoids the severe errors observed for SF-BLYP and is generally much less sensitive to the choice of functional, as is evident from Fig.~\ref{fig:st_gap}. It also yields better accuracy than SF-TDDFT for molecules dominated by essentially one-center diradical character  (Table~\ref{tb:st_gap}). At the BHHLYP level, for example, the MAE is 4.74 kcal/mol for TD$\Delta$SCF, compared with 11.67 kcal/mol for SF-TDDFT. Remarkably, quite similar MAEs are obtained with BLYP (4.82 kcal/mol) and B3LYP (4.41 kcal/mol), indicating that the overall performance of TD$\Delta$SCF is only weakly dependent on the functional. These results indicate that TD$\Delta$SCF  is more reliable than SF-TDDFT for the present set of singlet--triplet gaps. 

At the same time, however, TD$\Delta$SCF shows a clear systematic tendency to overestimate $\Delta E_{\rm ST}$ relative to experiment for all systems and functionals examined, in contrast to SF-TDDFT.  This implies that TD$\Delta$SCF systematically overestimates the singlet energy. This is likely because the lowest singlet state obtained by TD$\Delta$SCF depends strongly on the orbitals of the underlying $\Delta$SCF reference and is therefore not variationally optimized to the lowest-energy singlet solution, whereas the triplet state is fully optimized by a conventional open-shell SCF procedure. This systematic bias becomes particularly visible for $p$-benzyne, for which TD$\Delta$SCF predicts the wrong sign of $\Delta E_{\rm ST}$ with B3LYP and BHHLYP.

Although the functional dependence of TD$\Delta$SCF is weak in terms of the overall MAE, several notable exceptions are found at the level of individual systems, namely O$_2$, $o$-benzyne, and $p$-benzyne. For these systems, the variation among the three functionals is substantially larger than in the rest of the data set, where the discrepancy between functionals is less than $\sim$2 kcal/mol in most cases. At present, the origin of this behavior is not entirely clear. One possible explanation is that the underlying $\Delta$SCF reference states are not fully consistent across the different functionals, so that the subsequent TD$\Delta$SCF calculations are effectively based on slightly different electronic references. A more detailed analysis of the corresponding $\Delta$SCF states would be needed to clarify this point, but we leave such an analysis for future work.

\begin{figure}
    \centering
    \includegraphics[width=1\linewidth]{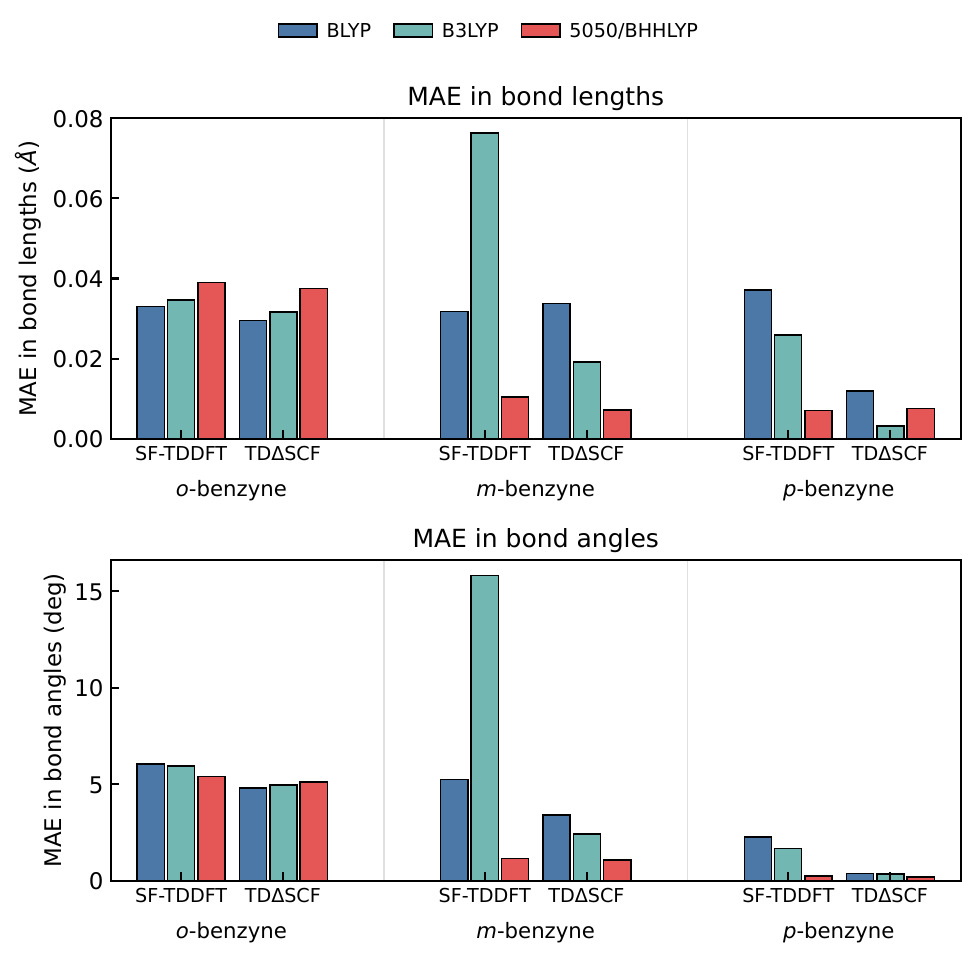}
    \caption{Mean absolute errors in bond lengths (\AA) and bond angles (deg) relative to the SF-CCSD reference.}
    \label{fig:geom2}
\end{figure}
Finally, the spin expectation values $\langle S^2 \rangle$ of the singlet states obtained by TD$\Delta$SCF are all below 0.1, indicating that spin contamination is negligible (see Supporting Information).

\subsection{Geometry optimization of Benzyne systems} 
Because TD$\Delta$SCF is formulated within the standard TDDFT/TDA framework, its analytical nuclear gradients are readily available.\cite{Furche02} We therefore carried out geometry optimizations of the singlet states  of the benzyne systems using TD$\Delta$SCF and SF-TDDFT. The 6-311G* basis set was used for both methods. In TD$\Delta$SCF, BLYP, B3LYP, and BHHLYP were employed as in the preceding sections, whereas in SF-TDDFT, the 50-50 functional was adopted instead of BHHLYP, as in Ref.~\cite{Bernard12}. The optimized bond lengths and bond angles are shown in Fig.~\ref{fig:geom}, and the corresponding MAEs relative to the SF-CCSD reference are summarized in Fig.~\ref{fig:geom2}. The SF-CCSD reference values were taken from Ref.~\cite{Bernard12}.

For $o$-benzyne and $p$-benzyne, there is no substantial difference between the geometries obtained with TD$\Delta$SCF and SF-TDDFT, regardless of the functional employed. Consistent with this observation, comparison with the SF-CCSD reference shows that the bond lengths are in good overall agreement for both systems, although some differences remain in the bond angles of $o$-benzyne. The MAEs relative to SF-CCSD are less than 0.04 {\AA} and approximately 6$^\circ$ for all methods as summarized in Fig. ~\ref{fig:geom2}.

In contrast, Fig.~\ref{fig:geom} and Fig.~\ref{fig:geom2} show that SF-B3LYP gives a markedly different geometry for $m$-benzyne, most notably by significantly underestimating the C1--C2--C3 angle relative to the other results. 
Specifically, this angle is 98.28$^{\circ}$ at the SF-CCSD level, whereas SF-B3LYP gives 71.83$^{\circ}$. In addition, the C1--C3 bond length is 2.0608~\AA\ at the SF-CCSD level  but 1.5835~\AA\ for SF-B3LYP, indicating that SF-B3LYP yields a structure in which C1 and C3 are much closer to each other. 
This qualitative difference is also reflected in the overall deviations from SF-CCSD: for $m$-benzyne, SF-B3LYP gives substantially larger MAEs in both bond lengths and bond angles than the TD$\Delta$SCF geometries and SF-5050. The full set of optimized geometrical parameters is provided in the Supporting Information.

In $m$-benzyne, such a shortening of the C1--C3 distance is known to be a geometrical feature of a bicyclic structure in which a bond is formed between C1 and C3.  Indeed, Kraka and co-workers reported that, depending on the theoretical method, a bicyclic structure can be stabilized for $m$-benzyne.\cite{Kraka} In contrast, Wenthold and co-workers provided photoelectron spectroscopic evidence supporting a non-bicyclic structure for m-benzyne,\cite{benzyne} and  Winkler and Sander reexamined the potential energy surface of $m$-benzyne and concluded that the minimum structure corresponds to a monocyclic geometry in which the C1--C3 distance remains around $2.05\pm 0.05$ {\AA}.\cite{Winkler} Furthermore, Al-Saidi and Umrigar showed by diffusion Monte Carlo calculations that the monocyclic structure is lower in energy than the bicyclic one.\cite{Al-Saidi}  Taken together, these results indicate that the C1--C3-contracted structure obtained by SF-B3LYP can be interpreted as a bicyclic structure and further suggest that SF-B3LYP fails to reproduce the monocyclic structure regarded as intrinsically stable for singlet $m$-benzyne, instead possibly over-stabilizing the bicyclic form. By contrast, TD$\Delta$SCF yields a monocyclic structure with a C1--C3 distance maintained around $2.09$--$2.20$ {\AA} for all functionals examined. These results indicate that TD$\Delta$SCF provides a more consistent monocyclic structural description for singlet $m$-benzyne than SF-TDDFT.

\begin{figure*}
    \centering
    \includegraphics[width=1\linewidth]{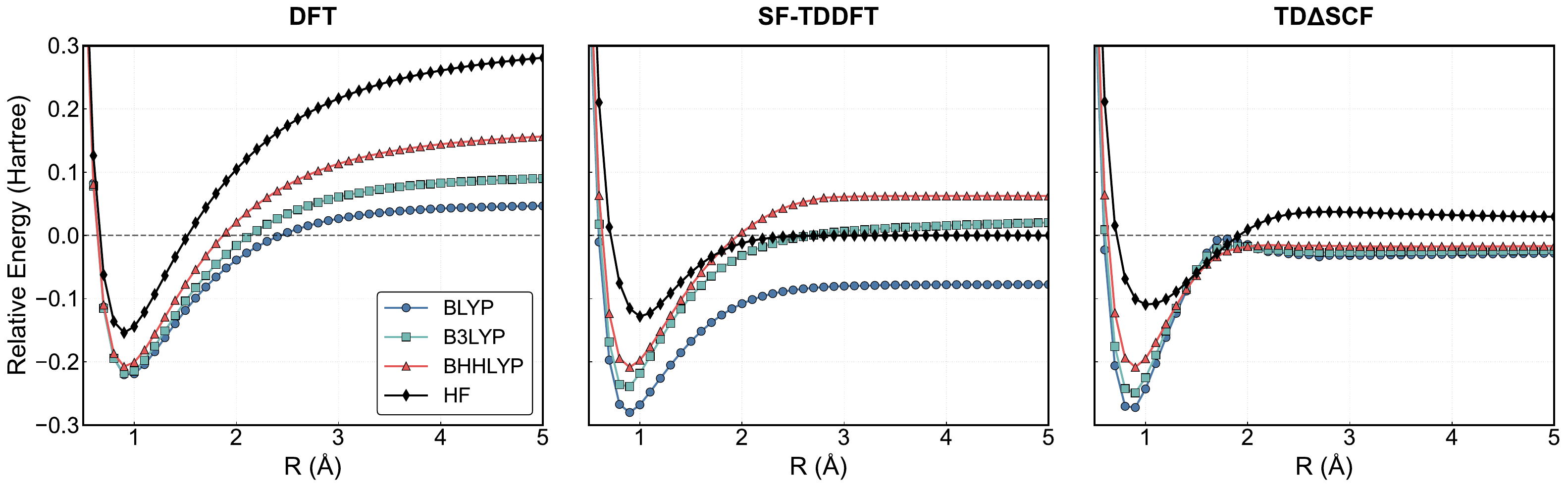}
    \caption{Bond dissociation curves of hydrogen fluoride calculated with each method. Energies are plotted relative to the gray dashed line, which denotes the unrestricted dissociation limit. The DFT curve was obtained from restricted Kohn-Sham calculations.}
    \label{fig:HF}
\end{figure*}

\begin{figure*}
    \centering
    \includegraphics[width=1\linewidth]{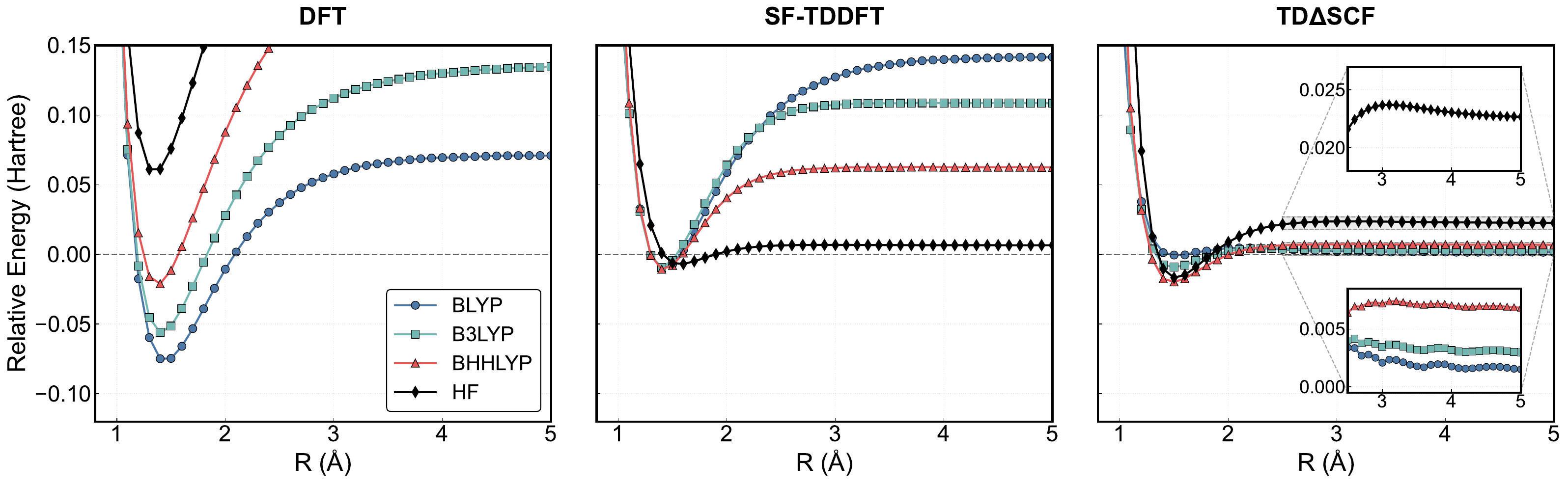}
    \caption{Bond dissociation curves of the fluorine molecule calculated with each method. Energies are plotted relative to the gray dashed line, which represents the dissociation limit obtained from unrestricted calculations. The DFT curve was obtained from restricted Kohn-Sham calculations.  Insets show enlarged views of the long-bond region.}
    \label{fig:F2}
\end{figure*}
\begin{table*}[t]
\centering
\caption{Bond dissociation energies $D_e$ (kcal/mol) of hydrogen fluoride and the fluorine molecule calculated with each method and functional.}
\label{tb:De}
\renewcommand{\arraystretch}{1.15}
\begin{threeparttable}
\begin{tabular}{c@{\hskip 10pt}c@{\hskip 10pt}c@{\hskip 10pt}c@{\hskip 10pt}c@{\hskip 6pt}c@{\hskip 6pt}c@{\hskip 10pt}c@{\hskip 10pt}c@{\hskip 10pt}c}
\hline\hline
\multirow{2}{*}{Method} & \multicolumn{4}{c}{Hydrogen fluoride} &  & \multicolumn{4}{c}{Fluorine molecule} \\ \cline{2-5} \cline{7-10}
& BLYP & B3LYP & BHHLYP & HF &  & BLYP & B3LYP & BHHLYP & HF \\ \hline
DFT            & 168.60 & 194.07 & 228.42 & 272.92 &  & 92.32 & 119.66 & 158.90 & 212.35 \\
SF-TDDFT       & 127.16 & 164.79 & 170.12 & 80.49  &  & 95.46 & 73.91 &46.01 & 8.53  \\
TD$\Delta$SCF  & 155.93 & 142.77 & 120.24 & 87.32  &  & 1.43  & 7.57   & 16.67 & 24.77 \\ \hline
Experiment     & \multicolumn{4}{c}{141.13$^{a}$} &  & \multicolumn{4}{c}{38.30$^{b}$} \\ \hline\hline
\end{tabular}
\begin{tablenotes}[flushleft]
\footnotesize
\item[$^{a}$] Ref.~\citenum{Zemke1991}.
\item[$^{b}$] Estimated from the reported value of $D_0$ in Ref.~\citenum{Wang2024}.
\end{tablenotes}
\end{threeparttable}
\end{table*}

\subsection{Potential energy curves}
Single-bond dissociation provides a stringent test for any method intended to describe near-degenerate electronic structures. In the stretched-bond region, the exact wave function acquires strong static-correlation character, and a balanced description of the bonding and antibonding configurations becomes essential. We therefore examine the bond-dissociation curves of hydrogen fluoride and the fluorine molecule (F$_2$) using DFT, SF-TDDFT, and TD$\Delta$SCF. This comparison is also of independent interest because, to the best of our knowledge, the performance of SF-TDDFT for such bond-dissociation curves has not been systematically analyzed. All calculations were carried out with the 6-31++G** basis set using BLYP, B3LYP, BHHLYP, and HF as the underlying functionals.

The resulting curves are shown in Figs.~\ref{fig:HF} and \ref{fig:F2}. In both cases, the energies are plotted relative to the dissociation limit obtained from unrestricted Kohn--Sham calculations, indicated by the gray dashed line. For both molecules, DFT gives a reasonable description near the equilibrium geometry. As the bond is stretched, however, the DFT curves do not approach this limit. This failure reflects the essentially single-determinantal nature of DFT: although the neutral singlet should become degenerate with the triplet reference in the proper dissociation limit, the underlying restricted description does not recover this degeneracy.

A major complication in SF-TDDFT is that, in the stretched-bond region, the target neutral bond-breaking singlet is often not the lowest spin-flip root. In Figs.~\ref{fig:HF} and \ref{fig:F2}, we therefore plot the singlet state associated with $\sigma$-bond breaking; the other low-lying SF-TDDFT states are omitted from the figures and summarized instead in the Supporting Information. For hydrogen fluoride, SF-TDDFT improves the asymptotic behavior relative to DFT only in limited cases. At the HF level, the dissociation curve approaches the unrestricted dissociation limit reasonably well. By contrast, with BLYP and B3LYP, severe self-interaction errors (SIEs) stabilize an unphysical ionic state, H$^+\cdots$F$^-$, below the neutral bond-breaking state. As a result, the physically relevant neutral dissociation state H$^\bullet\cdots$F$^\bullet$ lies at higher energy, remains strongly spin contaminated ($\langle \hat S^2\rangle \approx 1$), and does not cross the ionic branch (see Supporting Information). With BHHLYP, the SIE-driven ionic solution present with BLYP and B3LYP disappears, and the asymptotic region is instead governed by a neutral bond-breaking state. This state, however, still remains significantly spin contaminated and lies above the lowest roots, which correspond to $\pi\rightarrow \sigma^*$ excitations rather than to the target singlet. For the symmetric F$_2$ molecule, no analogous SIE-driven ionic solution is present, but the stretched-bond region still contains many low-lying spurious states. Overall, the SF-TDDFT dissociation curves exhibit strong functional dependence.

This pronounced functional dependence can be understood from the structure of the SF-TDDFT energy expression. The singlet energy is written as
\begin{equation}
E_{\rm S}^{\rm SF} = E_{\rm T} + {\bf X}^\dagger {\bf A}^{\rm SF} {\bf X},
\end{equation}
where $E_{\rm T}$ is the triplet reference energy and the second term is the spin-flip excitation energy. In the proper dissociation limit for these molecules, the triplet state becomes degenerate with the target singlet ground state, so the second term should approach zero. In collinear SF-TDDFT, however, this excitation energy generally remains finite and strongly functional dependent, because the spin-flip kernel lacks part of the usual Coulomb and exchange-correlation response. As a result, the required cancellation is incomplete, and static correlation is not recovered in a balanced manner, except in favorable cases such as SF-HF. This interpretation is also consistent with the ethylene torsion results discussed in Sec.~\ref{sec:ethylene}, where the description of the electronically degenerate $90^\circ$ geometry varied significantly among functionals and reasonable barrier heights were obtained only through error cancellation between different geometries.

TD$\Delta$SCF exhibits qualitatively different behavior from SF-TDDFT. First, the dissociation curves obtained with different density functionals are much more similar to one another, indicating markedly weaker functional dependence. More importantly, in the stretched-bond region, the TD$\Delta$SCF curves remain much closer to the unrestricted dissociation limit for both hydrogen fluoride and F$_2$, irrespective of the underlying functional, and no analogous spurious lower-lying states are observed. Thus, although TD$\Delta$SCF is not exact, it provides a substantially more balanced description of static correlation along bond dissociation.

This behavior can be understood from the corresponding energy expression,
\begin{equation}
E_{\rm S}^{\Delta \rm SCF} = E_{\Delta{\rm SCF}} + {\bf X}^\dagger {\bf A}^{\Delta{\rm SCF}} {\bf X}.
\end{equation}
The second term represents the response correction built on the broken-symmetry $\Delta$SCF reference and, within the adiabatic approximation, retains the usual exchange-correlation response contribution. In the dissociation limit, the $\Delta$SCF determinant may be regarded approximately as an equal mixture of the triplet state and the excited open-shell singlet, whose energies differ by about $2K_{\sigma\sigma^*}$, where $K_{\sigma\sigma^*}$ denotes the effective exchange interaction between the $\sigma$ and $\sigma^*$ orbitals. Therefore, one may write
\begin{equation}
E_{\Delta \rm SCF} \approx E_{\rm T} + K_{\sigma\sigma^*},
\end{equation}
suggesting that the response term compensates for this excess and drives the final TD$\Delta$SCF energy toward the neutral dissociation limit. In this sense, the de-excitation response in TD$\Delta$SCF plays a role analogous to an effective $-K_{\sigma\sigma^*}$ correction.

Another important distinction from SF-TDDFT is that TD$\Delta$SCF does not produce the spurious ionic H$^+\cdots$F$^-$ state as a competing low-lying root. This can also be understood from the structure of the TD$\Delta$SCF response space. In the dissociation limit, the bonding and antibonding orbitals may be expressed as
\begin{align}
|\sigma\rangle &= c |{\rm H}_{1s}\rangle + \sqrt{1-c^2}\,|{\rm F}_{2p_z}\rangle,\\
|\sigma^*\rangle &= \sqrt{1-c^2}\,|{\rm H}_{1s}\rangle - c |{\rm F}_{2p_z}\rangle,
\end{align}
so that
\begin{align}
|{\rm F}_{2p_z}\rangle = \sqrt{1-c^2}\,|\sigma\rangle - c |\sigma^*\rangle .
\end{align}
The ionic state can then be expanded as
\begin{align}
|{\rm H}^{+}\cdots{\rm F}^{-}\rangle
&=
|{\rm F}_{2p_z}^\alpha {\rm F}_{2p_z}^\beta\rangle \\
&=
(1-c^2)|\sigma^\alpha \sigma^\beta\rangle
+ c^2 |\sigma^{*\alpha} \sigma^{*\beta}\rangle
\notag\\
&\quad
- c\sqrt{1-c^2}
\left(
|\sigma^{*\alpha}\sigma^\beta\rangle
+
|\sigma^\alpha\sigma^{*\beta}\rangle
\right).
\end{align}
Since the reference $\Delta$SCF determinant is $|\sigma^{*\alpha}\sigma^\beta\rangle$, the closed-shell determinants $|\sigma^\alpha \sigma^\beta\rangle$ and $|\sigma^{*\alpha}\sigma^{*\beta}\rangle$ are reached by a single de-excitation and a single excitation, respectively, and can therefore appear within the linear-response space. By contrast, the determinant $|\sigma^\alpha\sigma^{*\beta}\rangle$ differs from the reference by two spin-orbitals and thus lies outside the singles response manifold. Consequently, a pure ionic state cannot be represented exactly within TD$\Delta$SCF built on $|\sigma^{*\alpha}\sigma^\beta\rangle$. This limitation is, in practice, beneficial because it prevents the spurious H$^+\cdots$F$^-$ solution from emerging as a low-lying state.

At the same time, however, TD$\Delta$SCF is not uniformly better over the entire potential energy curve. Near the equilibrium geometry, its performance depends more noticeably on the quality of the underlying $\Delta$SCF reference. With BHHLYP, the TD$\Delta$SCF curves remain reasonably close to the DFT curves in the bonding region. With BLYP and B3LYP, however, clear deviations emerge: for hydrogen fluoride the energy rises too steeply upon bond stretching, whereas for F$_2$ the minimum itself is placed too high. This behavior is naturally interpreted as an orbital bias inherited from the non-Aufbau reference determinant. Because the underlying orbitals are optimized for an excited-state determinant rather than for the final ground-state singlet, they are not guaranteed to provide a balanced description in the near-equilibrium region.

These trends are also reflected in the dissociation energies, $D_e$, summarized in Table~\ref{tb:De}. We estimate $D_e$ as the energy difference between the equilibrium geometry and the structure at $R=5$~\AA. The experimental values are 141.13 kcal/mol for hydrogen fluoride \cite{Zemke1991} and approximately 38.30 kcal/mol for F$_2$. The latter was estimated from the reported $D_0$ value of 37.00 kcal/mol,\cite{Wang2024}
where $D_0$ denotes the dissociation energy including the zero-point vibrational correction; the corresponding $D_e$ was obtained by adding this correction. 
For hydrogen fluoride, TD$\Delta$SCF reproduces the dissociation energy at least qualitatively for all functionals, and with B3LYP the agreement is even quantitatively reasonable. Although some functional dependence remains, it is much weaker than in SF-TDDFT. For F$_2$, in contrast, SF-TDDFT strongly overestimates $D_e$ except with BHHLYP, with errors of roughly 57 and 36 kcal/mol for BLYP and B3LYP, respectively. This directly reflects its failure to describe the dissociation limit properly. TD$\Delta$SCF, on the other hand, underestimates $D_e$ by about 22--37 kcal/mol. Importantly, this error does not mainly originate from the asymptotic region, which is described much better than in SF-TDDFT, but rather from the poorer description near the minimum, where the orbital bias of the $\Delta$SCF reference raises the energy. Thus, TD$\Delta$SCF is not uniformly accurate along the entire potential energy curve, but it captures the physically essential dissociation behavior more satisfactorily than SF-TDDFT and does so with substantially weaker functional dependence.

A final feature worth noting is that the TD$\Delta$SCF curves obtained with BLYP, B3LYP, and BHHLYP exhibit weak oscillations and, in some cases, small discontinuities in the stretched-bond region, as illustrated most clearly for F$_2$ in Fig.~\ref{fig:F2}. By contrast, the corresponding HF-based curve remains smooth. This contrast strongly suggests that the anomaly originates from the exchange-correlation potential $v_{\mathrm{xc}}$, rather than from the Hartree--Fock exchange contribution. We analyze the origin of this behavior in the next section.

\begin{figure*}
    \centering
    \includegraphics[width=1\linewidth]{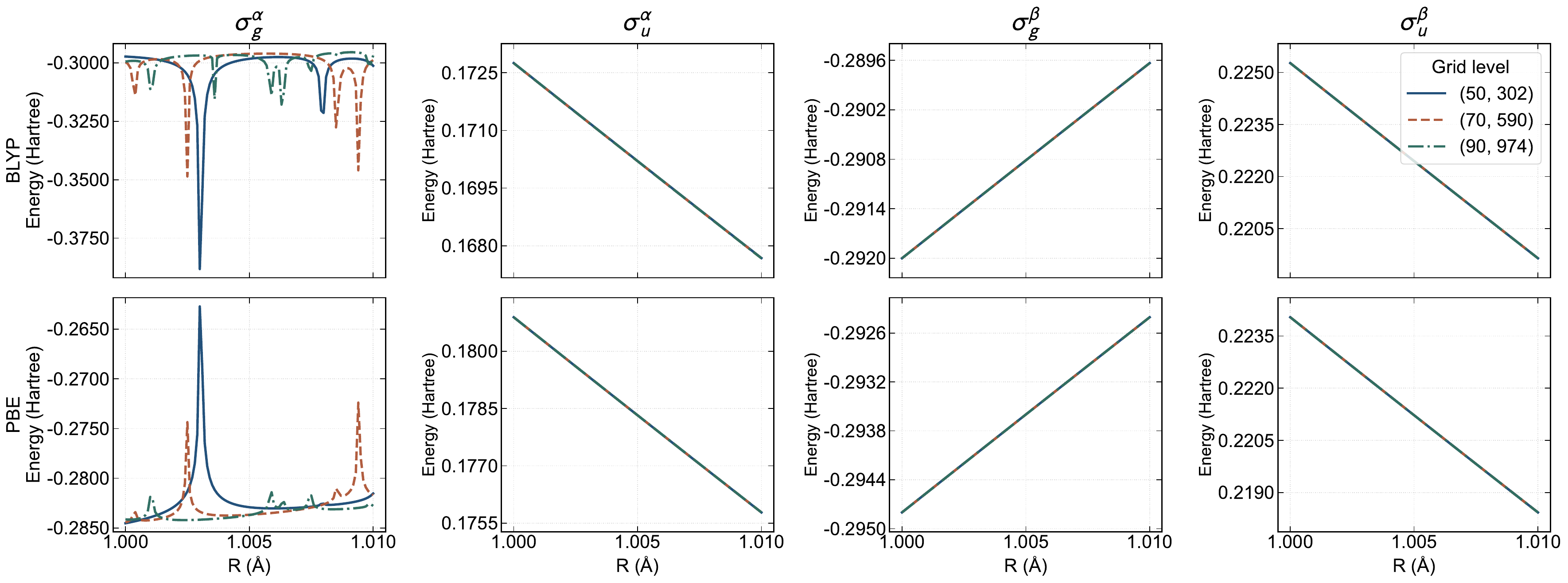}
    \caption{$\Delta$SCF orbital energies $\epsilon^{\Delta {\rm SCF}}$ of H$_2$ using BLYP (Top) and PBE (Bottom). $\sigma_u^\alpha$ and $\sigma_g^\beta$ are occupied, whereas $\sigma_g^\alpha$ and $\sigma_u^\beta$ are unoccupied.}
    \label{fig:orbital_energy}
\end{figure*}

\begin{figure}
    \centering
    \includegraphics[width=1\linewidth]{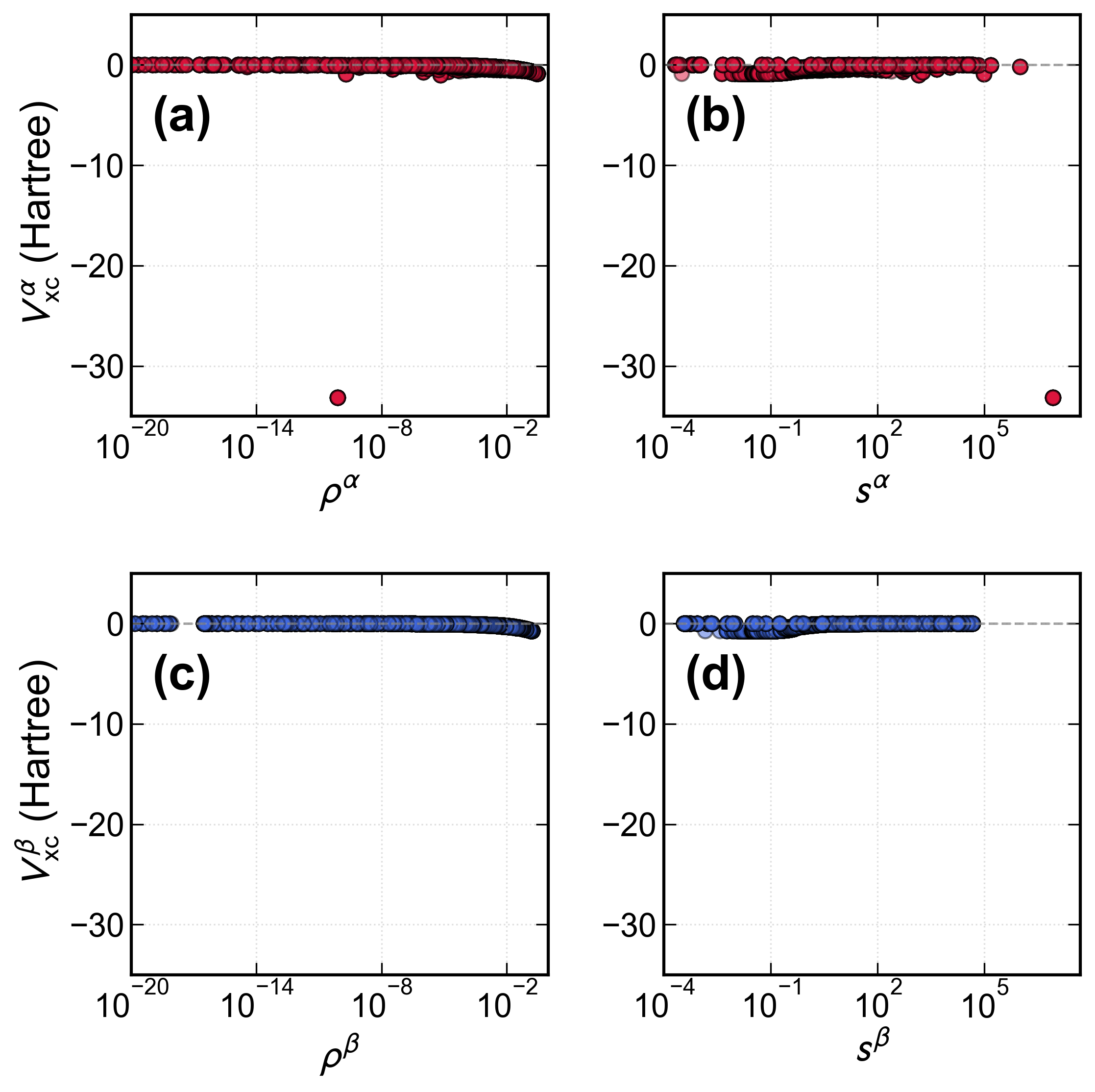}
    \caption{Scatter plots of the exchange--correlation potential $V_{\mathrm{xc}}(\mathbf r)$ against the electron density $\rho(\mathbf r)$ and the reduced gradient $s(\mathbf r)$, evaluated at all numerical grid points $\mathbf r$ for the $\Delta$SCF state of H$_2$ (see the main text). Panels (a) and (b) show the $\alpha$-spin components, whereas panels (c) and (d) show the $\beta$-spin components.}
    \label{fig:s_vs_Vxc}
\end{figure}

\begin{figure*}
    \centering
    \includegraphics[width=1\linewidth]{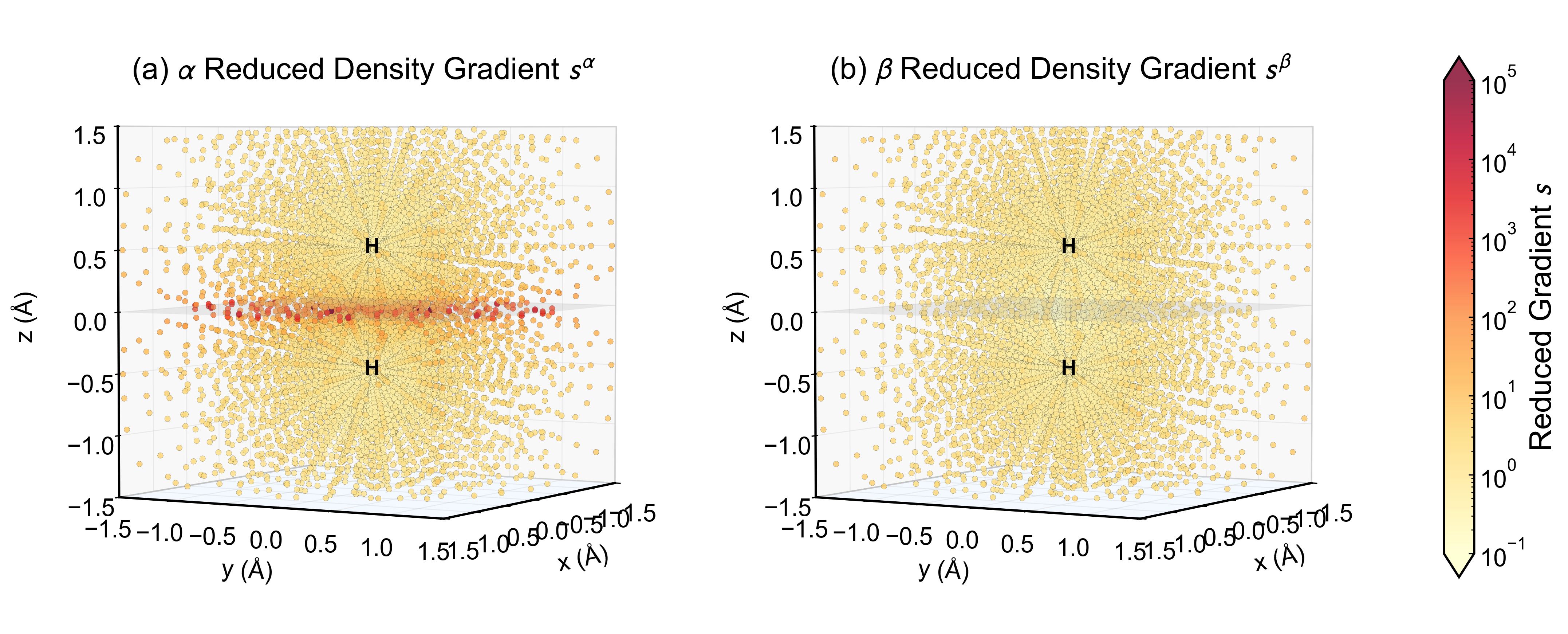}
    \caption{Three-dimensional distribution of the reduced density gradient $s(\mathbf r)$ at all numerical grid points for the $\Delta$SCF state of H$_2$. Panels (a) and (b) correspond to the $\alpha$- and $\beta$-spin components, respectively. The H--H bond is oriented along the $z$ axis, while the translucent blue slice marks the nodal plane of the antibonding orbital.}
    \label{fig:reduced_gradient}
\end{figure*}

\subsection{ Numerical Instability in Non-Aufbau Excitation: The Case of $\sigma_g \to \sigma_u$ Transition in H$_2$}\label{sec:grid}
In the course of our investigation of the potential energy curves in the previous section, we found that calculations of non-Aufbau electronic states with the $\Delta$SCF method can, under certain conditions, exhibit numerical instabilities that in turn lead to discontinuous behavior in TD$\Delta$SCF. To illustrate this in the simplest possible setting, we consider the potential energy curves of H$_2$ in the STO-3G basis, which contains only the  $\sigma_g$ and $\sigma_u$ orbitals. We impose a non-Aufbau occupation in which the $\alpha$ electron occupies $\sigma_u$ and the $\beta$ electron occupies $\sigma_g$. This minimal model removes complications arising from orbital relaxation and from discontinuities associated with changes in the SCF solution itself, thereby allowing us to isolate the behavior of the exchange-correlation potential $v_{\rm xc}$. Calculations were carried out with BLYP and PBE using three numerical integration grids, (50, 302), (70, 590), and  (90, 974).

Figure~\ref{fig:orbital_energy} shows the orbital energies of $\sigma_g$ and $\sigma_u$ for both spins as functions of bond length. Strikingly, the $\alpha$-spin $\sigma_g$ orbital energy exhibits sharp peaks. No comparable discontinuities appear in the other orbital energies, which behave smoothly as expected. We also find that both the positions and the magnitudes of these peaks depend on the numerical integration grid, whereas the peak positions are nearly identical for BLYP and PBE. Moreover, this pathological behavior is absent at the HF level (not shown). Taken together, these observations strongly suggest that the instability originates from the numerical evaluation of the exchange correlation potential $v_{\rm xc}({\bf r})$ in $\Delta$SCF. Because the KS orbital energies enter directly into the diagonal part of the TDDFT response matrix, such discontinuities immediately contaminate the TDDFT excitation energies. In addition, the TDDFT kernel itself is constructed from derivatives of the exchange-correlation potential and is therefore expected to inherit the same numerical pathology. The discontinuous behavior observed in the TD$\Delta$SCF energies is thus naturally understood as a consequence of instabilities in both the orbital-energy contribution and the exchange-correlation kernel. 

To examine the origin of this behavior more closely, Figure \ref{fig:s_vs_Vxc}(a) plots $v^\alpha_{\rm xc}({\bf r})$ against the $\alpha$-spin density $\rho^\alpha({\bf r})$ at $R=1.003$ {\AA} using the (50, 302) grid  for BLYP. A pronounced divergence in $v^\alpha_{\rm xc}$ appears at a particular grid point ${\bf r}^*$ where the density is extremely small, $\rho^\alpha({\bf r}^*) \approx 10^{-10}$. To identify the origin of this divergence more clearly, Figure~\ref{fig:s_vs_Vxc}(b) shows $v_{\rm xc}^\alpha$ as a function of the reduced gradient 
\begin{align}
    s({\bf r}) = \frac{|\nabla \rho({\bf r})|}{\rho^{4/3}({\bf r})}
\end{align}
The divergence in $v^\alpha_{\rm xc}$ is  clearly correlated with an extremely large value of $s({\bf r}^*)$, exceeding $10^6$. This indicates that $|\nabla \rho({\bf r}^*)|$ remains finite even though  $\rho({\bf r}^*)$ becomes vanishingly small, as can occur near a nodal surface.

This interpretation is confirmed by the three-dimensional analysis of $s$ in Figure \ref{fig:reduced_gradient}, which shows that the singularity is localized on the nodal plane ($z = 0$) of the antibonding orbital. Ordinarily, when the  electron density $\rho$ approaches zero at a point, its gradient $|\nabla\rho|$ is also expected to vanish. Here, however, the node arises from occupation of a single antibonding orbital, so  $\rho$ goes to zero while $|\nabla\rho|$ remains finite. This mismatch causes the reduced gradient $s$ to diverge locally, introducing a numerically singular contribution to the energy and potential that standard quadrature schemes cannot handle robustly. Indeed, the singular point ${\bf r}^*$ is found on the nodal plane. 

By contrast, the $\beta$-spin exchange-correlation potential remains well behaved, as shown in Figures \ref{fig:s_vs_Vxc}(c) and (d), and Figure~\ref{fig:reduced_gradient}(b). This is because the $\beta$ electron occupies the bonding $\sigma_g$ orbital, which has no node.

For ordinary ground-state densities, this type of instability is usually absent or negligibly small. In such cases, the nodal structure of an antibonding orbital is compensated by density contributions from other occupied orbitals, especially bonding orbitals, so that the total spin density remains finite even near the node. As a result, the reduced gradient $s({\bf r})$ does not diverge in the bonding region. The situation is qualitatively different near an uncompensated ``bare'' node, where no such compensating density is present. In that case, serious numerical and even physical breakdowns can arise because the density vanishes while its gradient remains finite. This problem becomes much more severe in dissociation limits or in localized electronic states, where the lack of compensation is most pronounced. In this sense, H$_2$ in the present setup should be regarded as a rather special, near-worst-case example, in which the instability is exposed in an especially clear and severe form. In the hydrogen fluoride and F$_2$ cases discussed in the previous section, by contrast, the same instability has only a minor effect on the overall shape of the potential energy curves. More generally, this issue is not expected to be problematic in many practical situations, including the ethylene torsion and singlet--triplet gaps discussed in Sections~\ref{sec:ethylene} and \ref{sec:st_gap}, because the antibonding density remains embedded in density contributions from other occupied orbitals, as in ordinary ground-state calculations. This interpretation is also consistent with the  hydrogen fluoride and F$_2$ potential energy curves, in which the discontinuity appears only in the dissociative region and has a negligible effect on the overall curve shape.

From this perspective, TD$\Delta$SCF may also be numerically more stable than noncollinear SF-TDDFT, whose kernel contains terms involving division by the spin-density difference $\rho_\alpha({\bf r}) - \rho_\beta({\bf r})$. Because regions in which the spin density becomes very small or vanishes are ubiquitous in molecular systems, such terms may represent an additional source of numerical instability.

\section{Concluding remarks}

In this work, we have introduced TD$\Delta$SCF, a linear-response formalism in which a non-Aufbau $\Delta$SCF determinant is used as the reference for a subsequent TDDFT/TDA calculation. In contrast to collinear SF-TDDFT, TD$\Delta$SCF retains the usual Coulomb and exchange-correlation response terms by employing a spin-conserving response manifold on top of an electronically promoted reference. The present results demonstrate that this alternative formulation provides a useful low-cost route for describing near-degenerate electronic structures.

For the ethylene torsion problem, TD$\Delta$SCF yields smooth potential-energy curves in the strongly near-degenerate region around the 90$^\circ$ twisted geometry for all functionals examined, whereas SF-BLYP exhibits an unphysical cusp. In the singlet--triplet gap benchmark, TD$\Delta$SCF is much less sensitive to the functional than collinear SF-TDDFT and gives consistently smaller overall errors, with MAEs of about 4.4--4.8 kcal/mol across BLYP, B3LYP, and BHHLYP. In addition, for the benzyne systems, TD$\Delta$SCF provides geometries that are generally comparable to those of SF-TDDFT for $o$- and $p$-benzyne and, importantly, yields a more consistent monocyclic description of singlet $m$-benzyne, whereas SF-B3LYP spuriously favors a bicyclic structure. For bond dissociation, TD$\Delta$SCF captures the dissociation behavior more satisfactorily than SF-TDDFT and shows substantially weaker functional dependence, although its accuracy near the equilibrium geometry is not uniformly improved.

At the same time, the present study also reveals several limitations of TD$\Delta$SCF. Although its overall functional dependence is weak, TD$\Delta$SCF shows a systematic tendency to overestimate singlet--triplet gaps, suggesting that the singlet states obtained from the response calculation are not fully optimized relative to the triplet states. Moreover, in bond dissociation, the method can lose accuracy near the equilibrium region when the underlying $\Delta$SCF reference is not appropriate, reflecting an orbital bias inherited from the non-Aufbau reference determinant. We also identify, for the first time to the best of our knowledge, a numerical instability that can arise in non-Aufbau DFT calculations and trace its origin to the exchange-correlation potential in the vicinity of uncompensated nodal regions.

Overall, TD$\Delta$SCF emerges as a promising single-determinant framework for singlet states with near-degenerate electronic structures, particularly in situations where the ground-state reference of ordinary TDDFT becomes qualitatively inadequate because of static correlation and collinear SF-TDDFT suffers from strong functional dependence. Further developments aimed at improving the construction and stability of the $\Delta$SCF reference, together with a deeper understanding of the numerical instabilities identified here, should broaden the applicability and reliability of this approach.

Several directions remain for future work. First, the present formulation uses a spin-mixed single-determinant non-Aufbau reference, and extensions based on spin-adapted or mixed-reference descriptions, including ROKS-type references and open-shell doublet references, would be valuable for reducing spin contamination and broadening the range of target states. Such developments would also be important for applications to conical intersections, where the relevant states must be described in a balanced manner within a common response space and the appropriate choice of the non-Aufbau reference is nontrivial. Second, it would be useful to assess TD$\Delta$SCF beyond the energetic and structural benchmarks considered here, including electronic-structure diagnostics such as natural-orbital occupations and diradical indices, as well as molecular properties such as dipole moments and spin--orbit couplings.

\section*{Associated content}
\subsection*{Supporting information}
Tables of the calculated $\Delta E_{\rm ST}$ values for each method and functional; $\langle\hat S^2\rangle$ values for the TD-$\Delta$SCF singlet states; bond-length and bond-angle deviations from SF-CCSD for $o$-, $m$-, and $p$-benzyne; TD-$\Delta$SCF dissociation curves of F$_2$ near the equilibrium bond distance and in the 2.5--5.0~\AA\ range; SF-TDDFT dissociation curves for the three lowest-energy states of hydrogen fluoride and the ten lowest-energy states of F$_2$; and tables of total energies and $\langle \hat S^2\rangle$ values for hydrogen fluoride and F$_2$.

\section*{Acknowledgments}
This work was supported by JSPS KAKENHI, Grant Nos. 25K01733 and 25K22247.

\section*{Notes}
The authors declare no competing financial interest.

\bibliography{refs}

@article{Bauernschmitt96,
title = {Treatment of electronic excitations within the adiabatic approximation of time dependent density functional theory},
journal = {Chemical Physics Letters},
volume = {256},
number = {4},
pages = {454-464},
year = {1996},
issn = {0009-2614},
doi = {https://doi.org/10.1016/0009-2614(96)00440-X},
url = {https://www.sciencedirect.com/science/article/pii/000926149600440X},
author = {Rüdiger Bauernschmitt and Reinhart Ahlrichs}
}

@book{ParrYang,
    author = {Parr, Robert G and Weitao, Yang},
    title = {Density-Functional Theory of Atoms and Molecules},
    publisher = {Oxford University Press},
    year = {1995},
    month = {01},
    isbn = {9780195092769},
    doi = {10.1093/oso/9780195092769.001.0001},
    url = {https://doi.org/10.1093/oso/9780195092769.001.0001}
}

@book{Ullrich,
    author = {Ullrich, Carsten A.},
    title = {Time-Dependent Density-Functional Theory: Concepts and Applications},
    publisher = {Oxford University Press},
    year = {2011},
    month = {12},
    isbn = {9780199563029},
    doi = {10.1093/acprof:oso/9780199563029.001.0001},
    url = {https://doi.org/10.1093/acprof:oso/9780199563029.001.0001}
}

@article{Petersilka96,
  title = {Excitation Energies from Time-Dependent Density-Functional Theory},
  author = {Petersilka, M. and Gossmann, U. J. and Gross, E. K. U.},
  journal = {Phys. Rev. Lett.},
  volume = {76},
  issue = {8},
  pages = {1212--1215},
  numpages = {0},
  year = {1996},
  month = {Feb},
  publisher = {American Physical Society},
  doi = {10.1103/PhysRevLett.76.1212},
  url = {https://link.aps.org/doi/10.1103/PhysRevLett.76.1212}
}

@article{CasidaHuixRotllant,
   author = "Casida, M.E. and Huix-Rotllant, M.",
   title = "Progress in Time-Dependent Density-Functional Theory", 
   journal= "Annual Review of Physical Chemistry",
   year = "2012",
   volume = "63",
   number = "Volume 63, 2012",
   pages = "287-323",
   doi = "https://doi.org/10.1146/annurev-physchem-032511-143803",
   url = "https://www.annualreviews.org/content/journals/10.1146/annurev-physchem-032511-143803",
   publisher = "Annual Reviews",
   issn = "1545-1593",
   type = "Journal Article"
  }

@article{Krylov06,
author = {Krylov, Anna I.},
title = {Spin-Flip Equation-of-Motion Coupled-Cluster Electronic Structure Method for a Description of Excited States, Bond Breaking, Diradicals, and Triradicals},
journal = {Accounts of Chemical Research},
volume = {39},
number = {2},
pages = {83-91},
year = {2006},
doi = {10.1021/ar0402006},
URL = { https://doi.org/10.1021/ar0402006
}}

@article{Minezawa09,
author = {Minezawa, Noriyuki and Gordon, Mark S.},
title = {Optimizing Conical Intersections by Spin-Flip Density Functional Theory: Application to Ethylene},
journal = {The Journal of Physical Chemistry A},
volume = {113},
number = {46},
pages = {12749-12753},
year = {2009},
doi = {10.1021/jp908032x},
    note ={PMID: 19905013},
URL = { https://doi.org/10.1021/jp908032x},
eprint = {https://doi.org/10.1021/jp908032x}
}

@article{Harabuchi14,
author = {Harabuchi, Yu and Keipert, Kristopher and Zahariev, Federico and Taketsugu, Tetsuya and Gordon, Mark S.},
title = {Dynamics Simulations with Spin-Flip Time-Dependent Density Functional Theory: Photoisomerization and Photocyclization Mechanisms of cis-Stilbene in $\pi\pi^\ast$ States},
journal = {The Journal of Physical Chemistry A},
volume = {118},
number = {51},
pages = {11987-11998},
year = {2014},
doi = {10.1021/jp5072428},
    note ={PMID: 25181251},
URL = { https://doi.org/10.1021/jp5072428},
eprint = { https://doi.org/10.1021/jp5072428}
}

@Article{Yue18,
author ="Yue, Ling and Liu, Yajun and Zhu, Chaoyuan",
title  ="Performance of TDDFT with and without spin-flip in trajectory surface hopping dynamics: cis–trans azobenzene photoisomerization",
journal  ="Phys. Chem. Chem. Phys.",
year  ="2018",
volume  ="20",
issue  ="37",
pages  ="24123-24139",
publisher  ="The Royal Society of Chemistry",
doi  ="10.1039/C8CP03851A",
url  ="http://dx.doi.org/10.1039/C8CP03851A"}

@article{Wang04,
    author = {Wang, Fan and Ziegler, Tom},
    title = {Time-dependent density functional theory based on a noncollinear formulation of the exchange-correlation potential},
    journal = {The Journal of Chemical Physics},
    volume = {121},
    number = {24},
    pages = {12191-12196},
    year = {2004},
    month = {12},
    issn = {0021-9606},
    doi = {10.1063/1.1821494},
    url = {https://doi.org/10.1063/1.1821494}
}

@article{Wang05,
    author = {Wang, Fan and Ziegler, Tom},
    title = {The performance of time-dependent density functional theory based on a noncollinear exchange-correlation potential in the calculations of excitation energies},
    journal = {The Journal of Chemical Physics},
    volume = {122},
    number = {7},
    pages = {074109},
    year = {2005},
    month = {02},
    issn = {0021-9606},
    doi = {10.1063/1.1844299},
    url = {https://doi.org/10.1063/1.1844299}
}

@article{Li12,
    author = {Li, Zhendong and Liu, Wenjian},
    title = {Theoretical and numerical assessments of spin-flip time-dependent density functional theory},
    journal = {The Journal of Chemical Physics},
    volume = {136},
    number = {2},
    pages = {024107},
    year = {2012},
    month = {01},
    issn = {0021-9606},
    doi = {10.1063/1.3676736},
    url = {https://doi.org/10.1063/1.3676736}
}

@article{Bernard12,
    author = {Bernard, Yves A. and Shao, Yihan and Krylov, Anna I.},
    title = {General formulation of spin-flip time-dependent density functional theory using non-collinear kernels: Theory, implementation, and benchmarks},
    journal = {The Journal of Chemical Physics},
    volume = {136},
    number = {20},
    pages = {204103},
    year = {2012},
    month = {05},
    issn = {0021-9606},
    doi = {10.1063/1.4714499},
    url = {https://doi.org/10.1063/1.4714499}
}

@article{Furche02,
    author = {Furche, Filipp and Ahlrichs, Reinhart},
    title = {Adiabatic time-dependent density functional methods for excited state properties},
    journal = {The Journal of Chemical Physics},
    volume = {117},
    number = {16},
    pages = {7433-7447},
    year = {2002},
    month = {10},
    issn = {0021-9606},
    doi = {10.1063/1.1508368},
    url = {https://doi.org/10.1063/1.1508368}
}

@article{Gilbert08,
  author = {Gilbert, A. T. and Besley, N. A. and Gill, P. M.},
  title = {Self-consistent field calculations of excited states using the maximum overlap method (MOM)},
  journal = {J. Phys. Chem. A},
  year = {2008},
  volume = {112},
  number = {50},
  pages = {13164--13171},
  doi = {10.1021/jp801738f}
}

@article{Carter-Fenk20,
author = {Carter-Fenk, Kevin and Herbert, John M.},
title = {State-Targeted Energy Projection: A Simple and Robust Approach to Orbital Relaxation of Non-Aufbau Self-Consistent Field Solutions},
journal = {Journal of Chemical Theory and Computation},
volume = {16},
number = {8},
pages = {5067-5082},
year = {2020},
doi = {10.1021/acs.jctc.0c00502},
    note ={PMID: 32644792},
URL = {https://doi.org/10.1021/acs.jctc.0c00502
},
eprint = {https://doi.org/10.1021/acs.jctc.0c00502}
}

@article{Hait20,
author = {Hait, Diptarka and Head-Gordon, Martin},
title = {Excited State Orbital Optimization via Minimizing the Square of the Gradient: General Approach and Application to Singly and Doubly Excited States via Density Functional Theory},
journal = {Journal of Chemical Theory and Computation},
volume = {16},
number = {3},
pages = {1699-1710},
year = {2020},
doi = {10.1021/acs.jctc.9b01127},
    note ={PMID: 32017554},
URL = {https://doi.org/10.1021/acs.jctc.9b01127},
eprint = {https://doi.org/10.1021/acs.jctc.9b01127}
}

@article{Filatov99,
    author = {Filatov, Michael and Shaik, Sason},
    title = {Application of spin-restricted open-shell Kohn–Sham method to atomic and molecular multiplet states},
    journal = {The Journal of Chemical Physics},
    volume = {110},
    number = {1},
    pages = {116-125},
    year = {1999},
    month = {01},
    issn = {0021-9606},
    doi = {10.1063/1.477941},
    url = {https://doi.org/10.1063/1.477941}
}

@article{Kowalczyk2013,
    author = {Kowalczyk, Tim and Tsuchimochi, Takashi and Chen, Po-Ta and Top, Laken and Van Voorhis, Troy},
    title = "{Excitation energies and Stokes shifts from a restricted open-shell Kohn-Sham approach}",
    journal = {J. Chem. Phys.},
    volume = {138}, 
    number = {16},
    pages = {164101},
    year = {2013},
    month = {04},
    issn = {0021-9606},
    doi = {10.1063/1.4801790},
    url = {https://doi.org/10.1063/1.4801790}
}

@Article{Levi20A,
author ="Levi, Gianluca and Ivanov, Aleksei V. and Jónsson, Hannes",
title  ="Variational calculations of excited states via direct optimization of the orbitals in DFT",
journal  ="Faraday Discuss.",
year  ="2020",
volume  ="224",
issue  ="0",
pages  ="448-466",
publisher  ="The Royal Society of Chemistry",
doi  ="10.1039/D0FD00064G",
url  ="http://dx.doi.org/10.1039/D0FD00064G"}

@article{Levi20B,
  author = {Levi, G. and Ivanov, A. V. and J$\'{o}$nsson, H.},
  title = {Variational Density Functional Calculations of Excited States via Direct Optimization},
  journal = {J. Chem. Theory Comput.},
  year = {2020},
  volume = {16},
  pages = {6968},
  doi = {10.1021/acs.jctc.0c00597}
}

@article{Hait21,
  author = {Hait, D. and Head-Gordon, M.},
  title = {Orbital Optimized Density Functional Theory for Electronic Excited States},
  journal = {J. Phys. Chem. Lett.},
  year = {2021},
  volume = {12},
  pages = {4517--4529},
  doi = {10.1021/acs.jpclett.1c00744}
}

@Article{Selenius2024,
author={Selenius, Elli
and Sigurdarson, Alec El{\'i}as
and Schmerwitz, Yorick L. A.
and Levi, Gianluca},
title={Orbital-Optimized Versus Time-Dependent Density Functional Calculations of Intramolecular Charge Transfer Excited States},
journal=JCTC,
year={2024},
month={May},
day={14},
publisher={American Chemical Society},
volume={20},
number={9},
pages={3809-3822},
issn={1549-9618},
doi={10.1021/acs.jctc.3c01319},
url={https://doi.org/10.1021/acs.jctc.3c01319}
}

@article{Hait20B,
author = {Hait, Diptarka and Head-Gordon, Martin},
title = {Highly Accurate Prediction of Core Spectra of Molecules at Density Functional Theory Cost: Attaining Sub-electronvolt Error from a Restricted Open-Shell Kohn–Sham Approach},
journal = {The Journal of Physical Chemistry Letters},
volume = {11},
number = {3},
pages = {775-786},
year = {2020},
doi = {10.1021/acs.jpclett.9b03661},
    note ={PMID: 31917579},
URL = { https://doi.org/10.1021/acs.jpclett.9b03661},
eprint = {https://doi.org/10.1021/acs.jpclett.9b03661}

}

@Article{Paetow25,
author ="Paetow, Lukas and Neugebauer, Johannes",
title  ="Excited state dipole moments from $\Delta$SCF: a benchmark",
journal  ="Phys. Chem. Chem. Phys.",
year  ="2025",
volume  ="27",
issue  ="31",
pages  ="16354-16370",
publisher  ="The Royal Society of Chemistry",
doi  ="10.1039/D5CP01695A",
url  ="http://dx.doi.org/10.1039/D5CP01695A"}

@article{Carter-Fenk22,
author = {Carter-Fenk, Kevin and Cunha, Leonardo A. and Arias-Martinez, Juan E. and Head-Gordon, Martin},
title = {Electron-Affinity Time-Dependent Density Functional Theory: Formalism and Applications to Core-Excited States},
journal = {The Journal of Physical Chemistry Letters},
volume = {13},
number = {41},
pages = {9664-9672},
year = {2022},
doi = {10.1021/acs.jpclett.2c02564},
    note ={PMID: 36215404},
URL = {https://doi.org/10.1021/acs.jpclett.2c02564},
eprint = {https://doi.org/10.1021/acs.jpclett.2c02564}

}

@article{Knepp25,
author = {Knepp, Zachary J. and Fertal, Domenica R. and Masso, Gabriel B. and Hamburger, Robert C. and Guzman, Christian A. and Young, Elizabeth R. and Fredin, Lisa A.},
title = {Predicting Excited-State Absorption Spectra from Non-Aufbau Configurations},
journal = {Journal of Chemical Theory and Computation},
volume = {21},
number = {19},
pages = {9736-9752},
year = {2025},
doi = {10.1021/acs.jctc.5c00591},
    note ={PMID: 41031514},
URL = {    https://doi.org/10.1021/acs.jctc.5c00591},
eprint = { https://doi.org/10.1021/acs.jctc.5c00591}

}

@Article{Berera09,
author={Berera, Rudi
and van Grondelle, Rienk
and Kennis, John T. M.},
title={Ultrafast transient absorption spectroscopy: principles and application to photosynthetic systems},
journal={Photosynthesis Research},
year={2009},
month={Sep},
day={01},
volume={101},
number={2},
pages={105-118},
issn={1573-5079},
doi={10.1007/s11120-009-9454-y},
url={https://doi.org/10.1007/s11120-009-9454-y}
}

@article{Tsuchimochi24,
	title = {Double configuration interaction singles: {Scalable} and size-intensive approach for orbital relaxation in excited states and bond-dissociation},
	volume = {161},
	issn = {0021-9606},
	url = {https://doi.org/10.1063/5.0243710},
	doi = {10.1063/5.0243710},
	number = {24},
	journal = {The Journal of Chemical Physics},
	author = {Tsuchimochi, Takashi},
	month = dec,
	year = {2024},
	pages = {241102}
}

@article{Tsuchimochi26A,
author = {Tsuchimochi, Takashi and Mokhtar, Benjamin},
title = {Leveraging Configuration Interaction Singles for Qualitative Descriptions of Ground and Excited States: State-Averaging, Linear-Response, and Spin-Projection},
journal = {Journal of Chemical Theory and Computation},
volume = {0},
number = {0},
pages = {null},
year = {0},
doi = {10.1021/acs.jctc.6c00182},
URL = { 
        https://doi.org/10.1021/acs.jctc.6c00182
},
eprint = { https://doi.org/10.1021/acs.jctc.6c00182
}
}

@article{Tsuchimochi26B,
author = {Tsuchimochi, Takashi},
title = {Analytical Nuclear Gradients for State-Averaged Configuration Interaction Singles Variants: Application to Conical Intersections},
journal = {Journal of Chemical Theory and Computation},
volume = {0},
number = {0},
pages = {null},
year = {0},
doi = {10.1021/acs.jctc.6c00308},
    note ={PMID: 42012079},
URL = { 
        https://doi.org/10.1021/acs.jctc.6c00308
},
eprint = { 
        https://doi.org/10.1021/acs.jctc.6c00308
}

}

@article{QChem,
    author = {Epifanovsky, Evgeny and Gilbert, Andrew T. B. and Feng, Xintian and Lee, Joonho and Mao, Yuezhi and Mardirossian, Narbe and Pokhilko, Pavel and White, Alec F. and Coons, Marc P. and Dempwolff, Adrian L. and Gan, Zhengting and Hait, Diptarka and Horn, Paul R. and Jacobson, Leif D. and Kaliman, Ilya and Kussmann, JÃ¶rg and Lange, Adrian W. and Lao, Ka Un and Levine, Daniel S. and Liu, Jie and McKenzie, Simon C. and Morrison, Adrian F. and Nanda, Kaushik D. and Plasser, Felix and Rehn, Dirk R. and Vidal, Marta L. and You, Zhi-Qiang and Zhu, Ying and Alam, Bushra and Albrecht, Benjamin J. and Aldossary, Abdulrahman and Alguire, Ethan and Andersen, Josefine H. and Athavale, Vishikh and Barton, Dennis and Begam, Khadiza and Behn, Andrew and Bellonzi, Nicole and Bernard, Yves A. and Berquist, Eric J. and Burton, Hugh G. A. and Carreras, Abel and Carter-Fenk, Kevin and Chakraborty, Romit and Chien, Alan D. and Closser, Kristina D. and Cofer-Shabica, Vale and Dasgupta, Saswata and de Wergifosse, Marc and Deng, Jia and Diedenhofen, Michael and Do, Hainam and Ehlert, Sebastian and Fang, Po-Tung and Fatehi, Shervin and Feng, Qingguo and Friedhoff, Triet and Gayvert, James and Ge, Qinghui and Gidofalvi, Gergely and Goldey, Matthew and Gomes, Joe and GonzÃ¡lez-Espinoza, Cristina E. and Gulania, Sahil and Gunina, Anastasia O. and Hanson-Heine, Magnus W. D. and Harbach, Phillip H. P. and Hauser, Andreas and Herbst, Michael F. and HernÃ¡ndez Vera, Mario and Hodecker, Manuel and Holden, Zachary C. and Houck, Shannon and Huang, Xunkun and Hui, Kerwin and Huynh, Bang C. and Ivanov, Maxim and \'{A}d\'{a}m J\'{a}sz and Ji, Hyunjun and Jiang, Hanjie and Kaduk, Benjamin and K\"{a}hler, Sven and Khistyaev, Kirill and Kim, Jaehoon and Kis, Gergely and Klunzinger, Phil and Koczor-Benda, Zsuzsanna and Koh, Joong Hoon and Kosenkov, Dimitri and Koulias, Laura and Kowalczyk, Tim and Krauter, Caroline M. and Kue, Karl and Kunitsa, Alexander and Kus, Thomas and Ladj\'{a}nszki, Istv\'{a}n and Landau, Arie and Lawler, Keith V. and Lefrancois, Daniel and Lehtola, Susi and Li, Run R. and Li, Yi-Pei and Liang, Jiashu and Liebenthal, Marcus and Lin, Hung-Hsuan and Lin, You-Sheng and Liu, Fenglai and Liu, Kuan-Yu and Loipersberger, Matthias and Luenser, Arne and Manjanath, Aaditya and Manohar, Prashant and Mansoor, Erum and Manzer, Sam F. and Mao, Shan-Ping and Marenich, Aleksandr V. and Markovich, Thomas and Mason, Stephen and Maurer, Simon A. and McLaughlin, Peter F. and Menger, Maximilian F. S. J. and Mewes, Jan-Michael and Mewes, Stefanie A. and Morgante, Pierpaolo and Mullinax, J. Wayne and Oosterbaan, Katherine J. and Paran, Garrette and Paul, Alexander C. and Paul, Suranjan K. and PavoÅ¡eviÄ Fabijan and Pei, Zheng and Prager, Stefan and Proynov, Emil I. and R\'{a}k, \'{A}d\'{a}m and Ramos-Cordoba, Eloy and Rana, Bhaskar and Rask, Alan E. and Rettig, Adam and Richard, Ryan M. and Rob, Fazle and Rossomme, Elliot and Scheele, Tarek and Scheurer, Maximilian and Schneider, Matthias and Sergueev, Nickolai and Sharada, Shaama M. and Skomorowski, Wojciech and Small, David W. and Stein, Christopher J. and Su, Yu-Chuan and Sundstrom, Eric J. and Tao, Zhen and Thirman, Jonathan and Tornai, GÃ¡bor J. and Tsuchimochi, Takashi and Tubman, Norm M. and Veccham, Srimukh Prasad and Vydrov, Oleg and Wenzel, Jan and Witte, Jon and Yamada, Atsushi and Yao, Kun and Yeganeh, Sina and Yost, Shane R. and Zech, Alexander and Zhang, Igor Ying and Zhang, Xing and Zhang, Yu and Zuev, Dmitry and Aspuru-Guzik, AlÃ¡n and Bell, Alexis T. and Besley, Nicholas A. and Bravaya, Ksenia B. and Brooks, Bernard R. and Casanova, David and Chai, Jeng-Da and Coriani, Sonia and Cramer, Christopher J. and Cserey, Gy\"orgy and DePrince, A. Eugene, III and DiStasio, Robert A., Jr. and Dreuw, Andreas and Dunietz, Barry D. and Furlani, Thomas R. and Goddard, William A., III and Hammes-Schiffer, Sharon and Head-Gordon, Teresa and Hehre, Warren J. and Hsu, Chao-Ping and Jagau, Thomas-C. and Jung, Yousung and Klamt, Andreas and Kong, Jing and Lambrecht, Daniel S. and Liang, WanZhen and Mayhall, Nicholas J. and McCurdy, C. William and Neaton, Jeffrey B. and Ochsenfeld, Christian and Parkhill, John A. and Peverati, Roberto and Rassolov, Vitaly A. and Shao, Yihan and Slipchenko, Lyudmila V. and Stauch, Tim and Steele, Ryan P. and Subotnik, Joseph E. and Thom, Alex J. W. and Tkatchenko, Alexandre and Truhlar, Donald G. and Van Voorhis, Troy and Wesolowski, Tomasz A. and Whaley, K. Birgitta and Woodcock, H. Lee, III and Zimmerman, Paul M. and Faraji, Shirin and Gill, Peter M. W. and Head-Gordon, Martin and Herbert, John M. and Krylov, Anna I.},
    title = {Software for the frontiers of quantum chemistry: An overview of developments in the Q-Chem 5 package},
    journal = {The Journal of Chemical Physics},
    volume = {155},
    number = {8},
    pages = {084801},
    year = {2021},
    month = {08},
    issn = {0021-9606},
    doi = {10.1063/5.0055522},
    url = {https://doi.org/10.1063/5.0055522}
}

@article{pyscf,
    author = {Sun, Qiming and Zhang, Xing and Banerjee, Samragni and Bao, Peng and Barbry, Marc and Blunt, Nick S. and Bogdanov, Nikolay A. and Booth, George H. and Chen, Jia and Cui, Zhi-Hao and Eriksen, Janus J. and Gao, Yang and Guo, Sheng and Hermann, Jan and Hermes, Matthew R. and Koh, Kevin and Koval, Peter and Lehtola, Susi and Li, Zhendong and Liu, Junzi and Mardirossian, Narbe and McClain, James D. and Motta, Mario and Mussard, Bastien and Pham, Hung Q. and Pulkin, Artem and Purwanto, Wirawan and Robinson, Paul J. and Ronca, Enrico and Sayfutyarova, Elvira R. and Scheurer, Maximilian and Schurkus, Henry F. and Smith, James E. T. and Sun, Chong and Sun, Shi-Ning and Upadhyay, Shiv and Wagner, Lucas K. and Wang, Xiao and White, Alec and Whitfield, James Daniel and Williamson, Mark J. and Wouters, Sebastian and Yang, Jun and Yu, Jason M. and Zhu, Tianyu and Berkelbach, Timothy C. and Sharma, Sandeep and Sokolov, Alexander Yu. and Chan, Garnet Kin-Lic},
    title = {Recent developments in the PySCF program package},
    journal = {The Journal of Chemical Physics},
    volume = {153},
    number = {2},
    pages = {024109},
    year = {2020},
    month = {07},
    issn = {0021-9606},
    doi = {10.1063/5.0006074},
    url = {https://doi.org/10.1063/5.0006074}
}

@article{Hirata99,
title = {Time-dependent density functional theory within the Tamm-Dancoff approximation},
journal = {Chemical Physics Letters},
volume = {314},
number = {3},
pages = {291-299},
year = {1999},
issn = {0009-2614},
doi = {https://doi.org/10.1016/S0009-2614(99)01149-5},
url = {https://www.sciencedirect.com/science/article/pii/S0009261499011495},
author = {So Hirata and Martin Head-Gordon}
}

@article{Nikiforov14,
    author = {Nikiforov, Alexander and Gamez, Jose A. and Thiel, Walter and Huix-Rotllant, Miquel and Filatov, Michael},
    title = {Assessment of approximate computational methods for conical intersections and branching plane vectors in organic molecules},
    journal = {The Journal of Chemical Physics},
    volume = {141},
    number = {12},
    pages = {124122},
    year = {2014},
    month = {09},
    issn = {0021-9606},
    doi = {10.1063/1.4896372},
    url = {https://doi.org/10.1063/1.4896372}
}

@article{Saade24,
author = {Saade, Sandra and Burton, Hugh G. A.},
title = {Excited State-Specific CASSCF Theory for the Torsion of Ethylene},
journal = {Journal of Chemical Theory and Computation},
volume = {20},
number = {12},
pages = {5105-5114},
year = {2024},
doi = {10.1021/acs.jctc.4c00212},
    note ={PMID: 38847452},
URL = { 
    
  https://doi.org/10.1021/acs.jctc.4c00212
    
    

}
}

@article{TDDFT,
  title = {Density-Functional Theory for Time-Dependent Systems},
  author = {Runge, Erich and Gross, E. K. U.},
  journal = {Phys. Rev. Lett.},
  volume = {52},
  issue = {12},
  pages = {997--1000},
  numpages = {0},
  year = {1984},
  month = {Mar},
  publisher = {American Physical Society},
  doi = {10.1103/PhysRevLett.52.997},
  url = {https://link.aps.org/doi/10.1103/PhysRevLett.52.997}
}

@article{Krylov01_B,
title = {Size-consistent wave functions for bond-breaking: the equation-of-motion spin-flip model},
journal = {Chemical Physics Letters},
volume = {338},
number = {4},
pages = {375-384},
year = {2001},
issn = {0009-2614},
doi = {https://doi.org/10.1016/S0009-2614(01)00287-1},
url = {https://www.sciencedirect.com/science/article/pii/S0009261401002871},
author = {Anna I. Krylov}
}

@article{Shao03,
    author = {Shao, Yihan and Head-Gordon, Martin and Krylov, Anna I.},
    title = {The spin–flip approach within time-dependent density functional theory: Theory and applications to diradicals},
    journal = {The Journal of Chemical Physics},
    volume = {118},
    number = {11},
    pages = {4807-4818},
    year = {2003},
    month = {03},
    issn = {0021-9606},
    doi = {10.1063/1.1545679},
    url = {https://doi.org/10.1063/1.1545679}
}

@article{Huix10,
author ="Huix-Rotllant, Miquel and Natarajan, Bhaarathi and Ipatov, Andrei and Muhavini Wawire, C. and Deutsch, Thierry and Casida, Mark E.",
title  ="Assessment of noncollinear spin-flip Tamm–Dancoff approximation time-dependent density-functional theory for the photochemical ring-opening of oxirane",
journal  ="Phys. Chem. Chem. Phys.",
year  ="2010",
volume  ="12",
issue  ="39",
pages  ="12811-12825",
publisher  ="The Royal Society of Chemistry",
doi  ="10.1039/C0CP00273A",
url  ="http://dx.doi.org/10.1039/C0CP00273A"}

@article{B3LYP,
    author = {Becke, Axel D.},
    title = {Density-functional thermochemistry. III. The role of exact exchange},
    journal = {The Journal of Chemical Physics},
    volume = {98},
    number = {7},
    pages = {5648-5652},
    year = {1993},
    month = {04},
    issn = {0021-9606},
    doi = {10.1063/1.464913},
    url = {https://doi.org/10.1063/1.464913}
}

@article{KohnSham65,
  author = {Kohn, Walter and Sham, Lu Jeu},
  title = {Self-Consistent Equations Including Exchange and Correlation Effects},
  journal = {Physical Review},
  volume = {140},
  number = {4A},
  pages = {A1133--A1138},
  year = {1965},
  doi = {10.1103/PhysRev.140.A1133}
}

@article{Kraka,
title = {m-Benzyne and bicyclo[3.1.0]hexatriene – which isomer is more stable? – a quantum chemical investigation},
journal = {Chemical Physics Letters},
volume = {348},
number = {1},
pages = {115-125},
year = {2001},
issn = {0009-2614},
doi = {https://doi.org/10.1016/S0009-2614(01)01049-1},
url = {https://www.sciencedirect.com/science/article/pii/S0009261401010491},
author = {Elfi Kraka and Josep Anglada and Angelica Hjerpe and Michael Filatov and Dieter Cremer}
}

@article{Winkler,
	title = {The {Structure} of meta-{Benzyne} {RevisitedA} {Close} {Look} into $\sigma$-{Bond} {Formation}},
	volume = {105},
	issn = {1089-5639},
	url = {https://doi.org/10.1021/jp012100c},
	doi = {10.1021/jp012100c},
	number = {45},
	journal = {The Journal of Physical Chemistry A},
	author = {Winkler, Michael and Sander, Wolfram},
	month = nov,
	year = {2001},
	pages = {10422--10432},
}

@article{Al-Saidi,
	title = {Fixed-node diffusion {Monte} {Carlo} study of the structures of m-benzyne},
	volume = {128},
	issn = {0021-9606},
	url = {https://doi.org/10.1063/1.2902979},
	doi = {10.1063/1.2902979},
	number = {15},
	journal = {The Journal of Chemical Physics},
	author = {Al-Saidi, W. A. and Umrigar, C. J.},
	month = apr,
	year = {2008},
	pages = {154324},
}

@book{atoms,
  author       = {Moore, Charlotte E.},
  title        = {Atomic Energy Levels},
  volume       = {1},
  publisher    = {U.S. Government Printing Office},
  address      = {Washington, DC},
  year         = {1949},
  series       = {National Bureau of Standards Circular No. 467}
}

@book{diatomic,
  author = {Huber, K. P. and Herzberg, G.},
  title = {Constants of Diatomic Molecules},
  year = {1979},
  publisher = {Van Nostrand Reinhold},
  address = {New York}
}

@article{benzyne,
	title = {Ultraviolet {Photoelectron} {Spectroscopy} of the o-, m-, and p-{Benzyne} {Negative} {Ions}. {Electron} {Affinities} and {Singlet}-{Triplet} {Splittings} for o-, m-, and p-{Benzyne}},
	volume = {120},
	issn = {0002-7863},
	url = {https://doi.org/10.1021/ja9803355},
	doi = {10.1021/ja9803355},
	number = {21},
	journal = {Journal of the American Chemical Society},
	author = {Wenthold, Paul G. and Squires, Robert R. and Lineberger, W. C.},
	month = jun,
	year = {1998},
	pages = {5279--5290},
}

@article{Zemke1991,
title = {Improved potential energy curves and dissociation energies for HF, DF and TF},
journal = {Chemical Physics Letters},
volume = {177},
number = {4},
pages = {412-418},
year = {1991},
issn = {0009-2614},
doi = {https://doi.org/10.1016/0009-2614(91)85076-9},
url = {https://www.sciencedirect.com/science/article/pii/0009261491850769},
author = {Warren T. Zemke and William C. Stwalley and John A. Coxon and Photos G. Hajigeorgiou}
}

@article{Wang2024,
	title = {Bond {Dissociation} {Energy} of {F2} with {Spectroscopic} {Accuracy} {Measured} {Using} {Predissociation} and {Threshold} {Fragment} {Yield} {Spectroscopy}},
	volume = {15},
	url = {https://doi.org/10.1021/acs.jpclett.4c03256},
	doi = {10.1021/acs.jpclett.4c03256},
	number = {51},
	journal = {The Journal of Physical Chemistry Letters},
	author = {Wang, Peng and Gong, Shiyan and Mo, Yuxiang},
	month = dec,
	year = {2024},
	pages = {12594--12600},
}

@article{MRSFTDDFT,
    author = {Lee, Seunghoon and Filatov, Michael and Lee, Sangyoub and Choi, Cheol Ho},
    title = {Eliminating spin-contamination of spin-flip time dependent density functional theory within linear response formalism by the use of zeroth-order mixed-reference (MR) reduced density matrix},
    journal = {The Journal of Chemical Physics},
    volume = {149},
    number = {10},
    pages = {104101},
    year = {2018},
    month = {09},
    issn = {0021-9606},
    doi = {10.1063/1.5044202},
    url = {https://doi.org/10.1063/1.5044202}
}

@article{Nooijen95,
    author = {Nooijen, Marcel and Bartlett, Rodney J.},
    title = {Description of core-excitation spectra by the open-shell electron-attachment equation-of-motion coupled cluster method},
    journal = {The Journal of Chemical Physics},
    volume = {102},
    number = {17},
    pages = {6735-6756},
    year = {1995},
    month = {05},
    issn = {0021-9606},
    doi = {10.1063/1.469147},
    url = {https://doi.org/10.1063/1.469147}
}

@Article{Casanova20,
author ="Casanova, David and Krylov, Anna I.",
title  ="Spin-flip methods in quantum chemistry",
journal  ="Phys. Chem. Chem. Phys.",
year  ="2020",
volume  ="22",
issue  ="8",
pages  ="4326-4342",
publisher  ="The Royal Society of Chemistry",
doi  ="10.1039/C9CP06507E",
url  ="http://dx.doi.org/10.1039/C9CP06507E"}

@article{CH2,
    author = {Jensen, Per and Bunker, P. R.},
    title = {The potential surface and stretching frequencies of $\tilde X$ 3B1 methylene (CH2) determined from experiment using the Morse oscillator-rigid bender internal dynamics Hamiltonian},
    volume = {89},
	issn = {0021-9606},
	url = {https://doi.org/10.1063/1.455184},
	doi = {10.1063/1.455184},
	number = {3},
	journal = {The Journal of Chemical Physics},
	author = {Jensen, Per and Bunker, P. R.},
	month = aug,
	year = {1988},
	pages = {1327--1332},
}

@article{NH2+,
	title = {Photoionization of the amidogen radical},
	volume = {83},
	issn = {0021-9606},
	url = {https://doi.org/10.1063/1.449045},
	doi = {10.1063/1.449045},
	number = {9},
	journal = {The Journal of Chemical Physics},
	author = {Gibson, S. T. and Greene, J. P. and Berkowitz, J.},
	month = nov,
	year = {1985},
	pages = {4319--4328},
}

@article{SiH2,
	title = {Photoionization mass spectrometric studies of {SiHn} (n=1–4)},
	volume = {86},
	issn = {0021-9606},
	url = {https://doi.org/10.1063/1.452213},
	doi = {10.1063/1.452213},
	number = {3},
	journal = {The Journal of Chemical Physics},
	author = {Berkowitz, J. and Greene, J. P. and Cho, H. and Ruščić, B.},
	month = feb,
	year = {1987},
	pages = {1235--1248},
}

@article{PH2+,
	title = {A photoionization study of {PH}: {PH2} revisited},
	volume = {90},
	issn = {0021-9606},
	url = {https://doi.org/10.1063/1.456522},
	doi = {10.1063/1.456522},
	number = {1},
	journal = {The Journal of Chemical Physics},
	author = {Berkowitz, J. and Cho, H.},
	month = jan,
	year = {1989},

	pages = {1--6},
}

\end{document}